\documentclass[sigconf, nonacm]{acmart}
\makeatletter
\def\@ACM@checkaffil{
    \if@ACM@instpresent\else
    \ClassWarningNoLine{\@classname}{No institution present for an affiliation}%
    \fi
    \if@ACM@citypresent\else
    \ClassWarningNoLine{\@classname}{No city present for an affiliation}%
    \fi
    \if@ACM@countrypresent\else
        \ClassWarningNoLine{\@classname}{No country present for an affiliation}%
    \fi
}
\makeatother

\usepackage{xspace}
\usepackage{graphicx}
\usepackage{subcaption}
\usepackage{enumitem}
\usepackage{booktabs}
\usepackage{multirow}
\usepackage[ruled,linesnumbered]{algorithm2e}

\def\Snospace~{\S{}}

\newcommand{\sys}{Curator\xspace}
\newcommand{\paraspace}{\vspace{0.05in}}
\newcommand{\parab}[1]{\paraspace\noindent{\bf #1} }

\title{\sys: Efficient Indexing for Multi-Tenant Vector Databases}
\author{Yicheng Jin}
\affiliation{
  \institution {Duke University}
}
\author{Yongji Wu}
\affiliation{
  \institution {Duke University}
}
\author{Wenjun Hu}
\affiliation{
  \institution {Yale University}
}
\author{Bruce M. Maggs}
\affiliation{
  \institution {Duke University \\ Emerald Innovations}
}
\author{Xiao Zhang}
\affiliation{
  \institution {Duke University}
}
\author{Danyang Zhuo}
\affiliation{
  \institution {Duke University}
}

\begin{document}

\begin{abstract}

Vector databases have emerged as key enablers for bridging intelligent applications with unstructured data, providing generic search and management support for embedding vectors extracted from the raw unstructured data.
As multiple data users can share the same database infrastructure, multi-tenancy support for vector databases is increasingly desirable. This hinges on an efficient \textit{filtered search} operation, i.e., only querying the vectors accessible to a particular tenant.
Multi-tenancy in vector databases is currently achieved by building either a single, shared index among all tenants, or a per-tenant index. The former optimizes for memory efficiency at the expense of search performance, while the latter does the opposite.
Instead, this paper presents \sys, an in-memory vector index design tailored for multi-tenant queries that simultaneously achieves the two conflicting goals, low memory overhead and high performance for queries, vector insertion, and deletion.
\sys indexes each tenant's vectors with a tenant-specific clustering tree and encodes these trees compactly as sub-trees of a shared clustering tree. Each tenant's clustering tree adapts dynamically to its unique vector distribution, while maintaining a low per-tenant memory footprint.
Our evaluation, based on two widely used data sets, confirms that \sys delivers search performance on par with per-tenant indexing, while maintaining memory consumption at the same level as metadata filtering on a single, shared index. 

\end{abstract}

\maketitle

\pagestyle{plain}

\section{Introduction}

Vector similarity search has become a fundamental building block of wide-ranging intelligent applications, including search engines~\cite{Nayak_2019,Zeng_2022}, recommendation systems~\cite{zhang2019deep,naumov2019deep,covington2016deep}, and retrieval-based AI chatbots~\cite{chatgpt_retrieval_plugin,lewis2020retrieval}. This rise is driven by two factors. The first is the recent surge in volume and diversity of unstructured data, such as images, videos, and sensor logs, from sources like social media, e-commerce, streaming services, and IoT devices. According to IDC's estimate, 80\% of the world's data will be unstructured by 2025~\cite{King_2021}. At the same time, advancement in deep learning has made it possible to encode the semantics of such data into high-dimensional embedding vectors~\cite{le2014distributed,radford2021learning,narayanan2017graph2vec}, facilitating complex semantic analyses and information retrieval on unstructured data. 

The need to query and manage vector data has naturally led to the development of vector databases~\cite{weaviate,wang2021milvus,qdrant}. Due to the inherent difficulty of exact vector similarity search in high-dimensional spaces, vector databases usually opt for approximate nearest neighbor (ANN) search using specialized data structures known as vector indexes~\cite{johnson2019faiss,malkov2018hnsw}. The main advantage of this approach is that it can greatly improve search performance, with only a minor degradation in query accuracy.

\begin{figure}[t]
  \centering
  \includegraphics[width=0.9\columnwidth]{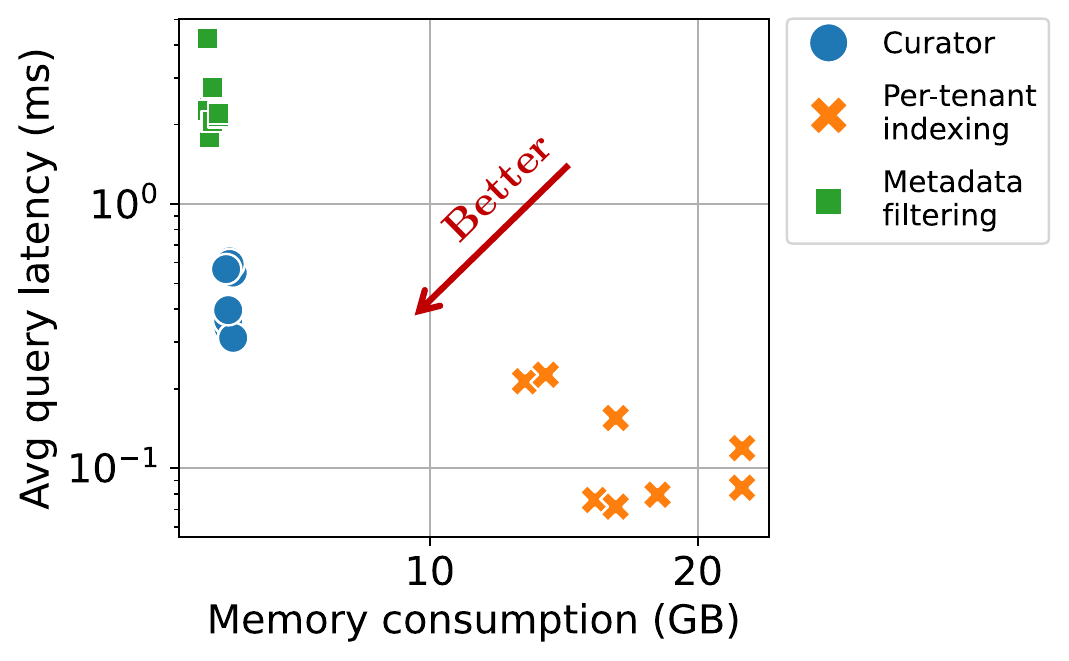}
  \vspace{-2mm}
  \caption{Current multi-tenancy strategies incur inherent inefficiency in either query latency or memory usage, while \sys aims to optimize both simultaneously.}
  \vspace{-5mm}
  \label{fig:comparison}
\end{figure}

As vector databases become indispensable software infrastructure, multi-tenancy support is increasingly needed.
For example, in a shared enterprise knowledge base with a semantic search engine, each document is converted into an embedding vector reflecting its semantics, which is recomputed whenever the document is updated. Document access within the knowledge base is contingent on the specific roles of users --- certain documents may be accessible to specific departments or levels of management, while others are more widely available. Furthermore, the system provides flexible sharing options, enabling users to dynamically grant access to documents to other users as needed, which is particularly useful for collaborative projects or cross-departmental interactions. This role-based, flexible access control is key to maintaining both security and functionality in the enterprise environment. This means when a user issues a vector similarity search in the database, the search result should only return vectors that a user has access to. Several leading vector databases, such as Pinecone~\cite{pinecone}, Milvus~\cite{wang2021milvus}, Weaviate~\cite{weaviate} and Qdrant~\cite{qdrant}, have recently announced support for multi-tenancy.

Currently, multi-tenancy support in vector databases is predominantly implemented via two strategies: (1) per-tenant indexing and (2) metadata filtering on a single, shared index. Per-tenant indexing creates individual indexes for each tenant in the database. This approach enhances search performance by enabling unfiltered queries specific to each tenant and adjusting the index structure to match the vector distribution particular to each tenant. However, this approach can significantly increase memory usage due to the overhead of per-tenant indexes and duplication of vector data among tenants when data sharing exists (detailed in~\ref{sec:background:multitenancy}). Conversely, metadata filtering uses a shared index with access lists attached to each vector as metadata. While this method is more memory-efficient and offers flexible access control, it can compromise search performance due to runtime permission checks and the mismatch between the shared index structure and tenant-specific vector distribution. Fig.~\ref{fig:comparison} illustrates this trade-off by comparing per-tenant indexing and metadata filtering in terms of search speed and memory usage to manage vectors generated by using the CLIP embedding model~\cite{radford2021learning} on the YFCC100M dataset. Each indexing approach generates multiple data points in  Fig.~\ref{fig:comparison} because we explore different combinations of configurations (in terms of the vector index's parameters) using parameter sweep. \autoref{sec:exp_setup} contains details hardware description of this experiment. On average, per-tenant indexing is $37.2\times$ faster than metadata filtering but has to consume $10.1\times$ more memory.

In this paper, we argue that this trade-off is not fundamental. The inherent limitations of the two existing approaches are due to them ``patching'' single-tenant indexes to support multi-tenant vector databases. Instead, if we design from first principle an index for multi-tenant vector databases, the index can achieve high search performance and low memory overhead at the same time.

To achieve this, we must address the key requirements and challenges below. (1) An efficient vector index must adapt its structure to accommodate the unique vector distribution of each tenant. (2) The index should minimize memory overhead by eliminating redundant data and minimizing the size of tenant-specific data structures. (3) The index must facilitate fast tenant-specific queries. In contrast to the metadata filtering strategy, the search algorithm must efficiently narrow the search scope to vectors accessible to the querying tenant.

\sys addresses these challenges with the following design. (1) It indexes each tenant's vectors with a hierarchical k-means clustering tree, whose structure dynamically adapts to the tenant's unique vector distribution. (2) The clustering trees for each tenant are compactly encoded with Bloom filters as sub-trees of a shared clustering tree, thereby minimizing per-tenant memory footprint. (3) \sys stores the access information of vectors in the index in a new format called shortlists, which facilitates fast search with no additional memory overhead. Our evaluation based on the YFCC100M and arXiv datasets shows that, \sys delivers search performance on par with per-tenant indexing, while maintaining a memory footprint at the same level as metadata filtering.

This paper makes the following contributions:
\begin{itemize}
    \item We design and implement \sys, an in-memory multi-tenant vector index that simultaneously achieves high search performance and a low memory footprint.
    \item We introduce a novel index structure that captures the unique data distributions of individual tenants with minimal memory overhead.
    \item We evaluate \sys on real-world text and image vector datasets. Our results confirm that \sys successfully delivers search performance on par with per-tenant indexing, while maintaining memory consumption at the same level as metadata filtering.
\end{itemize}

\section{Background}
\label{sec:background}

We first cover the common vector indexes used to accelerate vector similarity search. We then discuss the multi-tenancy strategies employed by existing vector databases and how their limitations motivate the design of \sys.

\subsection{Vector Indexes}

\textit{Vector similarity search} is fundamental to numerous modern applications, including search engines~\cite{Nayak_2019,Zeng_2022}, recommendation systems~\cite{zhang2019deep,naumov2019deep,covington2016deep}, and retrieval-augmented generation with large language models~\cite{chatgpt_retrieval_plugin,lewis2020retrieval}. Vector similarity search aims to find the nearest neighbors to a given point in a high-dimensional vector space. The curse of dimensionality, however, makes efficient exact similarity search inherently infeasible in terms of search time.
Consequently, real-world vector databases typically opt for \textit{approximate $k$-nearest neighbor search} ($k$-ANN search), which compromises search quality slightly in favor of higher search speed. A vector index is a core component in vector databases to support efficient ANN search.

The vector indexes that support ANN and that are currently widely adopted can be categorized into two families: graph-based and partition-based. Graph-based vector indexes represent vectors as vertices in a proximity graph, where each edge represents the neighboring relationship between the vectors it connects. At query time, the search algorithm efficiently traverses this graph towards the neighborhood of the query vector. Among many variants in this category, HNSW (Hierarchical Navigable Small World graph) index~\cite{malkov2018hnsw} is one of the most popular solutions. HNSW search starts from a fixed entry point and iteratively moves to the neighbor that is closer to the query vector. This greedy routing process continues until it reaches a local minima, when no neighbor is closer to the query vector. In order to find the top-k nearest neighbors to the query vector, the search algorithm then performs a best-first graph search starting from the local minima, and maintains a running queue of the top-$k$ candidates in this process. In HNSW, the length of the search path scales logarithmically with the size of the graph and the average degree of each node is bounded by a constant, thus allowing fast queries with logarithmic complexity. Storing the proximity graph, however, can significantly increase memory usage, making it less scalable to massive datasets~\cite{fu2019nsg}. 

Partition-based vector indexes, on the other hand, divide the vector space into sub-spaces using such techniques as clustering, locality sensitive hashing and space partitioning trees. During a search, these indexes first reduce the search scope to the sub-spaces closest to the query vector, followed by scanning vectors within these sub-spaces. The IVF (Inverted File) index~\cite{johnson2019faiss} is a widely-adopted member of this family, which uses clustering to partition the vector space and represents the vectors within each cluster by its centroid. The distance between a query vector and a cluster is determined by the distance to its centroid, which helps the search algorithm to focus on the most promising clusters. Although IVF does not offer the same logarithmic search complexity as HNSW, it still achieves reasonable search performance in practice. Moreover, IVF is usually much more memory efficient than HNSW, because it only needs to store the centroid of each cluster, instead of an edge list for each vector.

\subsection{Multi-Tenancy Support in Vector Databases}
\label{sec:background:multitenancy}

A multi-tenant database accommodates diverse clients within a shared database infrastructure to reduce the total cost of ownership, while ensuring minimal performance degradation and strict data isolation.
In current vector databases, there are two common strategies for implementing multi-tenancy: (1) per-tenant indexing and (2) metadata filtering.
\textbf{Per-tenant indexing} creates a separate index for each tenant within the same database instance and simply routes search queries from each tenant to its corresponding index at runtime. One key strength of this approach is its simplicity, because it does not alter the index structure or search algorithm.
On the other hand, the \textbf{metadata filtering} strategy stores all vectors in a shared index and associates each vector with an access list containing the IDs of the tenants with access rights to the vector. During search, queries are transformed to incorporate binary predicates that filter out vectors by checking if the querying tenant is included in the access list. 

In the metadata filtering strategy, there are various ways to execute predicated queries in vector indexes, depending on when the predicate filter is applied in the search process.
In \textit{pre-filtering}, filtering happens before vector search. It can either pre-partition the index offline according to attribute values and perform regular index scan on relevant partitions during searches~\cite{gollapudi2023filtered,wang2021milvus}, or it can construct a bitmap at runtime using traditional attribute filtering techniques and filter out vectors during the scan based on the bitmap~\cite{wei2020analyticdb,wang2021milvus}. Despite the simplicity of this approach, offline pre-filtering assumes predicates to be known in advance, and constructing bitmaps at runtime requires evaluating every vector against the predicate.
\textit{Post-filtering} first performs a regular index scan to generate an intermediate candidate set that contains closest neighbors to the query vector and then filters out inaccessible vectors~\cite{wei2020analyticdb,wang2021milvus}. For example, to compute top-$k$ with the filter, we can first use traditional ANN algorithms to find the top $2k$ vectors (the size of the candidate set is $2k$), and then use the filter to find the top-$k$ which pass the filter. Determining an appropriate size for the candidate set can be challenging, however, as the filter selectivity is typically unknown beforehand. If the size is too large, the search will become excessively costly; conversely, if it is too small, the result set may contain fewer items than expected by the query.
In contrast to the approaches above, \textit{single-stage filtering} integrates attribute information into the scan operator and evaluates each visited vector against the predicate~\cite{wu2022hqann,gollapudi2023filtered}. This approach eliminates the need for exhaustive permission checking as seen in pre-filtering, and can adaptively stop index scanning once a sufficient number of vectors have been found. Single-stage filtering is best known method for metadata filtering and is adopted by major vector databases already~\cite{pinecone,weaviate,qdrant}. 
Therefore, we use single-stage filtering as our evaluation baseline for metadata filtering. 

\begin{figure}[t]
  \centering
  \includegraphics[width=\columnwidth]{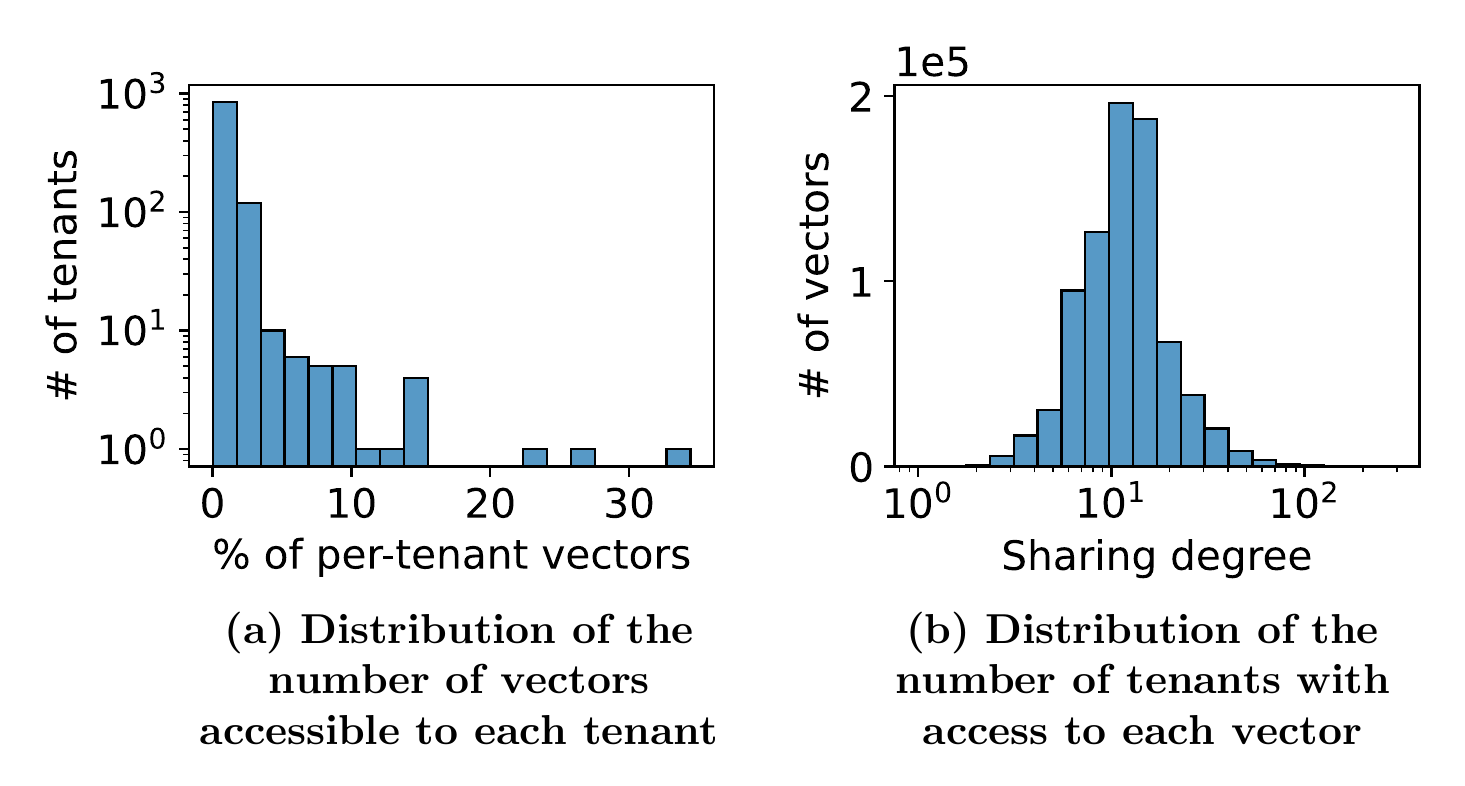}
  \vspace{-6mm}
  \caption{Statistics of the YFCC100M dataset. (a) Most tenants have access to only a small portion of all vectors. (b) On average, each vector is accessible to around 10 tenants, although some vectors are shared among up to 100 tenants.}
  \vspace{-5mm}
  \label{fig:dataset_stat}
\end{figure}

\parab{Trade-off between search speed and memory footprint.}
Unfortunately, these multi-tenancy strategies present a difficult trade-off for developers between search performance and memory usage. Per-tenant indexing offers better search performance, because each query only involves an unfiltered search within the querying tenant's index. Creating many small indexes, however, can result in high memory usage when vectors can be shared among tenants. The reasons for this are twofold: (1) The redundancy among the metadata stored in per-tenant indexes, e.g., cluster centroids in IVF and neighborhood graphs in HNSW. Take HNSW for example: If each vector is shared by 5 tenants and connects to 10 neighbors in each per-tenant index, we would need to store 50 edges per vector in total, a $5\times$ increase in index size compared to storing a single index. (2) The overhead of duplicated vector data across per-tenant indexes. Taking the same example when each vector is shared by 5 tenants, this means we need 5 total copies of vector data stored in the index for per-tenant indexing. 
Both problems come from the fact that data sharing is common. Similar sharing patterns exist in file systems, traditional multi-tenant databases~\cite{aulbach2011extensibility}, as well as the applications involving vector similarity search for unstructured data, such as recommendation systems. 
Empirically, we observe significant data sharing in our evaluation datasets: As shown in Fig.~\ref{fig:dataset_stat}b, most vectors in the YFCC100M dataset (detailed in \autoref{sec:exp_setup}) are shared by at least 10 tenants. In a multi-tenant environment where memory resources are shared, such memory overhead can significantly limit the degree of consolidation and diminish the benefits of economy of scale.

\begin{figure}[t]
  \centering
  \includegraphics[width=\columnwidth]{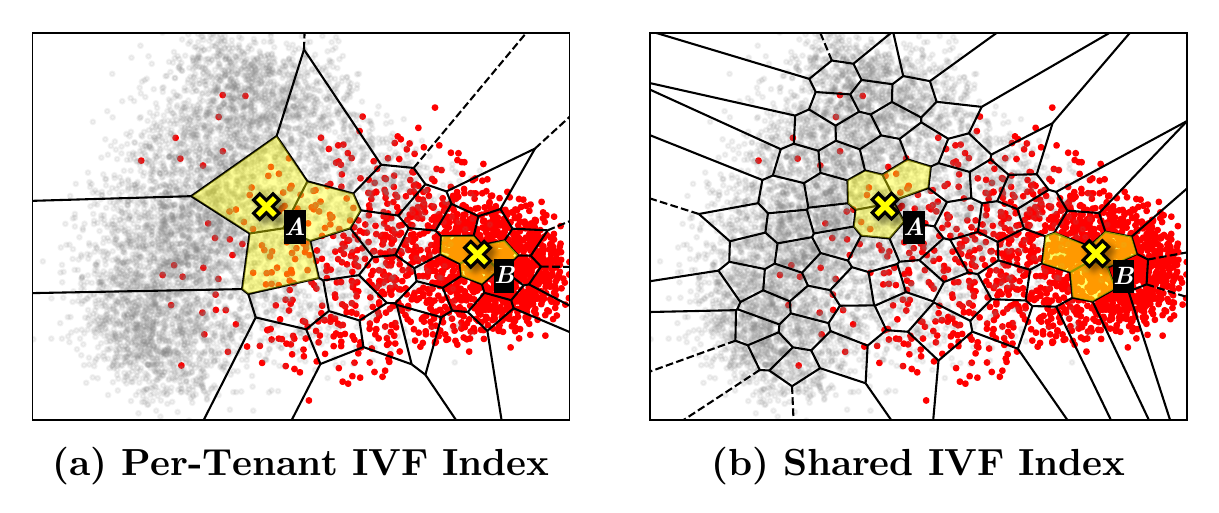}
  \vspace{-5mm}
  \caption{Shared index fails to capture the unique vector distribution of individual tenants, compromising search efficiency. 
  Vectors of the querying tenant are highlighted in red, while others are in gray.}
  \vspace{-5mm}
  \label{fig:motivation}
\end{figure}

Metadata filtering, on the other hand, introduces low memory overhead. By storing all vectors in a shared index, it avoids the overhead caused by redundancy as seen in per-tenant indexes. The only information it stores to implement multi-tenancy is the access list associated with each vector, which can be stored compactly in memory.  The search performance of metadata filtering, however, can be negatively affected by the following two aspects. (1) The overhead of runtime permission checking can significantly slow down search, especially when the filter is highly selective, i.e., each tenant only has access to a small portion of vectors. For instance, in the YFCC100M dataset (Fig.~\ref{fig:dataset_stat}a), most tenants can access less than 5\% of all vectors.
In this case, most vectors visited in the search process are irrelevant to the querying tenant, and checking each of them against the filter predicate can consume a significant portion of the search time. (2) Compared to per-tenant indexing, the structure of a shared index is not optimized for the unique vector distribution of individual tenants. Intuitively, a vector index should be constructed according to the underlying vector distribution, and a mismatch between the index structure and the vector distribution will undermine its efficiency.

To illustrate the latter point, we compare how a shared index with metadata filtering supports $k$-ANN search against per-tenant indexing in Fig.~\ref{fig:motivation}. For this motivation example, we transform the arXiv dataset into vectors using the \texttt{all-MiniLM-L6-v2}~\cite{huggingface_allMiniLM-L6-v2} embedding model and reduce them to 2D using PCA. Readers can easily observe the clear discrepancy between the overall vector distribution (gray points) and the querying tenant's vector distribution (red points): The vectors accessible to the querying tenant form a distinct cluster, while the overall vector distribution is evenly spread across a broader region. In contrast to the shared index (Fig~\ref{fig:motivation}a) that is constructed based on the overall vector distribution, the tenant-specific index (Fig~\ref{fig:motivation}b) adapts to the unique vector distribution of the querying tenant by partitioning the dense regions with finer granularity. This allows the search to explore different regions in the vector space with adaptive resolution. For example, assuming both indexes scan the three closest clusters to the query vector, the per-tenant index allows query A to explore a broader area with coarse-grained partitioning, while the shared index may fail to find $k$ vectors due to limited search scope. When executing query B in a dense region, the per-tenant index focuses search in a smaller neighborhood with higher resolution, thereby reducing the number of distance computations.

We attribute the inefficiency of both multi-tenancy strategies to the fact that neither the index structure nor the search algorithm is optimized for multi-tenant scenarios. As a result, we aim to design a vector index with native support for multi-tenancy that achieves the following goals simultaneously: (1) achieve similar search performance as per-tenant indexing, which involves adapting index structure to the unique vector distribution of each tenant and avoiding the overhead of metadata filtering; (2) achieve similar memory consumption as a single shared index, which requires minimizing data redundancy and arranging the access metadata in a way that allows fast search with minimal memory overhead.

\section{\sys Design Overview}
\label{sec:overview}

\begin{figure}[t]
  \centering
  \includegraphics[width=\columnwidth]{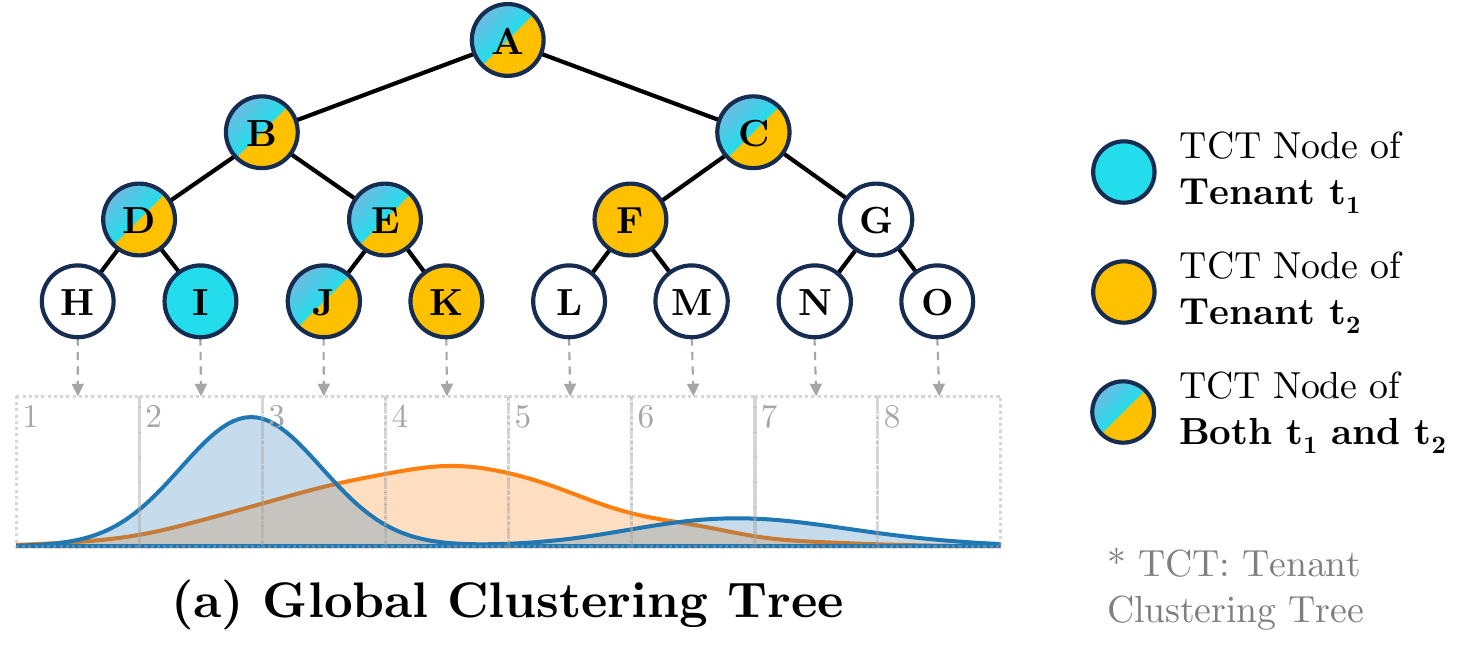}
  \includegraphics[width=\columnwidth]{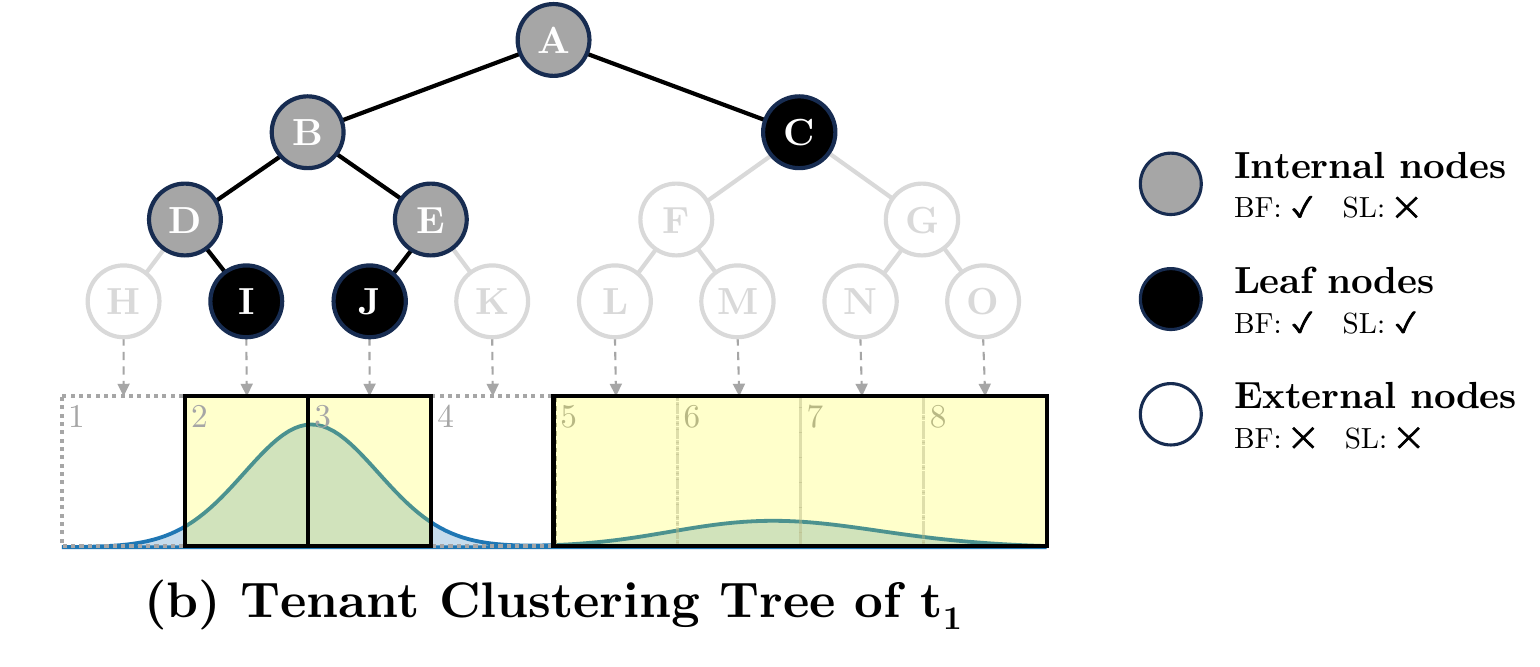}
  \includegraphics[width=\columnwidth]{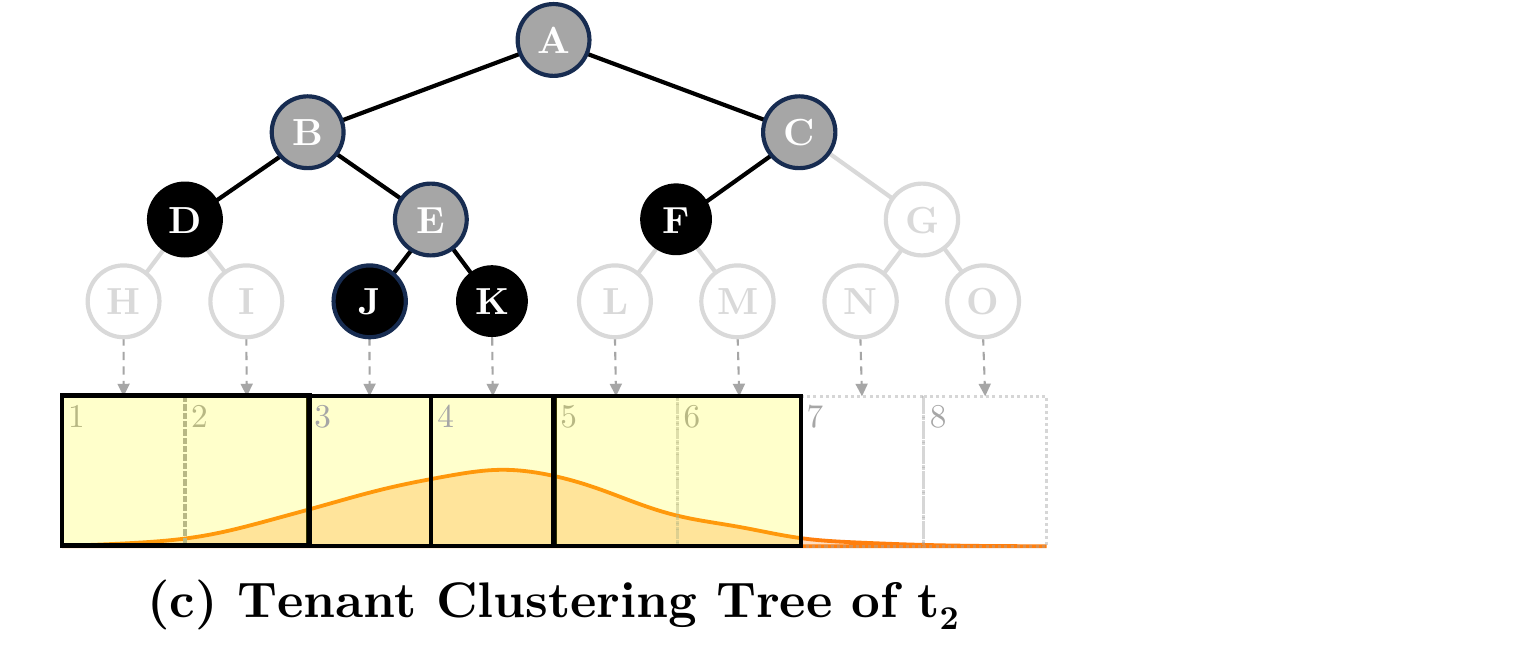}
  \vspace{-5mm}
  \caption{Overview of \sys. For simplicity, we represent the high-dimensional vector space as a line and restrict the branching factor of each tree node to 2. For a tree node $n$ in the Tenant Clustering Tree of tenant $t$, \texttt{BF} and \texttt{SL} in the legend stand for whether the Bloom filter of $n$ contains $t$ and whether $n$ contains a shortlist for $t$, respectively.}
 \vspace{-5mm}
  \label{fig:overview}
\end{figure}

Our high-level idea is the following: we would like to have a single vector index that stores each tenant's access information and at the same time have each tenant's view of the index be dependent on the data distribution of the tenant. If we can store the access information in a succinct matter, we can achieve small memory footprint (as does today's metadata filtering approach). We make the index dependent on each tenant's data distribution which makes search query efficient(as does today's per-tenant indexing approach~\cite{chiu2019learning,shrivastava2014asymmetric,bentley1975multidimensional,omohundro1989five,dasgupta2013randomized}).

\sys is composed of a shared Global Clustering Tree (Fig.~\ref{fig:overview}a) and per-tenant Tenant Clustering Trees (Fig.~\ref{fig:overview}b and~\ref{fig:overview}c) that encode each tenant's access control information in a compact fashion. The Global Clustering Tree, abbreviated as GCT, is a hierarchical k-means clustering tree that partitions vector space with multiple levels of granularity. This partitioning is with respect to the overall vector distribution and shared by all tenants. The Tenant Clustering Trees, abbreviated as TCTs, are sub-trees of GCT that index the vectors accessible to each tenant and adapt their structures to tenant-specific vector distributions.

As shown in Fig.~\ref{fig:overview}a, the GCT recursively bisects the vector space, illustrated in 1D for simplicity, into 8 non-overlapping clusters, labeled Clusters 1 to 8. Each leaf node in the GCT represents one of these clusters, while each internal node represents the union of the clusters represented by its child nodes. For example, node C represents the union of Clusters 5 through 8. 
Fig.~\ref{fig:overview}a also shows the TCTs of two tenants, $t_1$ and $t_2$, as well as the probability density functions (PDF) of their vector distributions. The TCT of each tenant shares the same tree structure as the GCT, but partitions the vector space differently: They grow deeper in regions where the corresponding tenant's vectors are densely distributed, thus creating a more fine-grained partition in these regions. For instance, as shown in Fig.~\ref{fig:overview}b, the vectors of $t_1$ exhibit a typical bimodal distribution. The left peak at Cluster 2-3, where vectors are densely concentrated, is divided by the TCT to the finest level of granularity, with nodes I and J corresponding to the two leaf clusters. For the right peak, with lower probability density, the TCT stops partitioning at level 1, grouping all vectors within a single cluster represented by node C. Note that nodes H and K are excluded from the TCT, because their corresponding clusters do not contain any vector from $t_1$. Overall, the leaf nodes of the TCT for tenant $t$ gives rise to a tenant-specific partition of the vector space (excluding the empty clusters), which we refer to as the implied IVF index of $t$, denoted $IVF(t)$.

Now, the remaining questions are (1) how to compactly encode the structure of the TCTs so as to reduce the per-tenant memory footprints and (2) how to avoid the overhead of permission checking in a mostly-shared index. Regarding the first question, since the TCTs share the same tree structure as the GCT, we only need to keep track of the nodes that belong to each TCT. To do so, we attach a \textit{Bloom filter}, a highly space-efficient data structure, to each GCT node to record the set of tenants whose TCT includes that node. For each TCT, the Bloom filters of its internal and leaf nodes should contain the corresponding tenant, while external nodes should not. To address the second question, we need to rethink how access information is maintained in the index. Rather than storing an access list for each vector, which necessitates expensive list traversal during permission checking, we store a list of vectors accessible to a specific tenant at each leaf node of the corresponding TCT. These lists, which we refer to as \textit{shortlists}, are pre-computed filtering results for each cluster of $IVF(t)$. It is worth noting that storing access information in shortlists does not increase memory usage, as it essentially represents the same access matrix in a different layout.
For example, in Fig.~\ref{fig:overview}a, the Bloom filter at node B contains $t_1$ and $t_2$ because it is shared by both TCTs. However, node B is not associated with any shortlist because it is not a leaf node in either TCT. On the other hand, node F's Bloom filter contains only $t_2$, and node F also contains a shortlist for $t_2$ that contains all vectors accessible to $t_2$ within Clusters 5 and 6. 

As shown in Fig.~\ref{fig:overview}b and~\ref{fig:overview}c, Bloom filters and shortlists together define the TCTs of tenant $t_1$ and $t_2$. For each tenant $t$, the GCT is divided into three regions. A node with a shortlist for $t$ must be a leaf node (shown in black) in the TCT. Otherwise, we examine the attached Bloom filter. If it contains tenant $t$, then the node is an internal node of the TCT (shown in gray). If not, the node must be outside the TCT, which we refer to as an external node (grayed out in the figure for contrast).

As depicted in Fig.~\ref{fig:overview}b and~\ref{fig:overview}c, the TCTs of tenant $t_1$ and $t_2$ are defined by Bloom filters and shortlists. For each tenant $t$, the GCT is divided into three regions by these filters and shortlists. A node with a shortlist for $t$ must be a leaf node (shown in black) in the TCT. If it's not, we examine the attached Bloom filter. If the filter contains tenant $t$, then the node is an internal node of the TCT. If it doesn't, the node is omitted from the TCT and is considered an external node.

\sys addresses the limitations inherent in existing multi-tenancy strategies as follows. Unlike the metadata filtering strategy, \sys avoids the overhead of permission checking by reorganizing access lists into shortlists, which allows the search algorithm to efficiently determine the subset of accessible vectors. Moreover, the structure of each tenant's TCT is aligned with its unique vector distribution for higher search efficiency. Compared to the per-tenant indexing strategy, \sys stores the TCT of each tenant in a space-efficient way, thereby significantly reducing memory usage.

\begin{table}[t]
\centering
\caption{Notations used in the paper}
\vspace{-3mm}
\label{tab:terms}
\small
\begin{tabular}{cll}
\toprule
& \textbf{Notation} & \textbf{Description} \\
\midrule
\multirow{4}{*}{\rotatebox[origin=c]{90}{\textbf{General}}}
& \( v \in V \) & Set of vectors \\
& \( t \in T \) & Set of tenants \\
& \( V(t) \) & Accessible vectors of \( t \) \\
& \( T(v) \) & Access list of \( v \) \\
\midrule
\multirow{9}{*}{\rotatebox[origin=c]{90}{\textbf{\sys-Specific}}}
& \( GCT \) & Global clustering tree \\
& \( n \in N \) & Nodes of GCT \\
& \( Desc(n) \) & Descendants of \( n \) \\
& \( V(n) \) & Subset of vectors within \( n \) \\
& \( V(n, t) = V(n) \cap V(t) \) & Vectors in \( n \) accessible to \( t \) \\
& \( leaf(v) \) & Assigned leaf node of \( v \) \\
& \( SL(n, t) \) & Shortlist for \( t \) at \( n \) \\
& \( SL(n) = \{t \in T \mid SL(n, t) \neq \varnothing\} \) & Tenants with shortlists at \( n \) \\
& \( BF(n) = \bigcup_{m \in Desc(n)} SL(m) \) & Bloom filter at \( n \) \\
& \( TCT(t) = \{n \in N \mid t \in BF(n) \} \) & Tenant clustering tree of \( t \) \\
\bottomrule
\end{tabular}
\vspace{-3mm}
\end{table}

\section{Details of \sys}
\label{sec:algorithms}

We first describe how similarity search works in \sys. We then describe how the index is constructed and how the index is maintained during data update (i.e., vector insertion/deletion, changes in the tenant access permissions).

\subsection{Similarity Search}
\label{sec:algorithm:search}

\begin{algorithm}[tb]

\SetAlgoLined
\SetAlgoNoEnd
\DontPrintSemicolon

\SetKw{And}{and}
\SetKw{Break}{break}
\SetKw{Continue}{continue}
\SetKwFunction{Distance}{Distance}\SetKwFunction{TopK}{TopK}
\SetKwInOut{Input}{Input}\SetKwInOut{Output}{Output}

\caption{$k$-ANN Search Algorithm}
\label{algo:hier_ivf_search}
\Input{query vector $x$, number of nearest neighbors to return $k$, tenant $t$, search radius $\gamma_1$, $\gamma_2$}
\Output{approximate top-$k$ nearest neighbors of $x$}
\BlankLine

\textcolor{blue}{\tcc{Stage 1. Identify candidate clusters with best-first search}}
$frontier \leftarrow \{(GCT.root, 0)\}$ \tcp*[r]{Search frontier}
$candClus \leftarrow \varnothing$ \tcp*[r]{Candidate clusters}
$nVecs \leftarrow 0$ \tcp*[r]{Total size of shortlists}

\While{$frontier \ne \varnothing$ \And $nCand < \gamma_1 \cdot \gamma_2 \cdot k$}{
    $(node, d) \leftarrow$ nearest element to $x$ in $frontier$\;
    \If(\tcp*[f]{Case 1}){$t \notin BF(node)$}{
        \Continue\;
    }
    \ElseIf(\tcp*[f]{Case 2}){$t \in SL(node)$}{
        $candClus \leftarrow candClus \cup (node, d)$\;
        $nVecs \leftarrow nVecs + |SL(node, t)|$\;
    }
    \Else(\tcp*[f]{Case 3}){
        \ForEach{$child \in node.children$}{
            $d \leftarrow \Distance(child.centroid, x)$\;
            $frontier \leftarrow frontier \cup (child, d)$\;
        }
    }
}
\BlankLine

\textcolor{blue}{\tcc{Stage 2. Scanning shortlists in candidate clusters in distance order}}
$candVecs \leftarrow \varnothing$ \tcp*[r]{Candidate vectors}
\While{$candClus \ne \varnothing$}{
    $(node, d) \leftarrow$ nearest cluster to $x$ in $candClus$\;
    \ForEach{$v \in SL(node, t)$}{
        $candVecs \leftarrow candVecs \cup (v, \Distance(v, x))$\;
    }
    \If{$|candVecs| \ge \gamma_1 \cdot k$}{
        \Break\;
    }
}
\BlankLine

\Return{$\TopK(candVecs, k)$}\;
\end{algorithm}

\begin{figure}[t]
  \centering
  \includegraphics[width=\columnwidth]{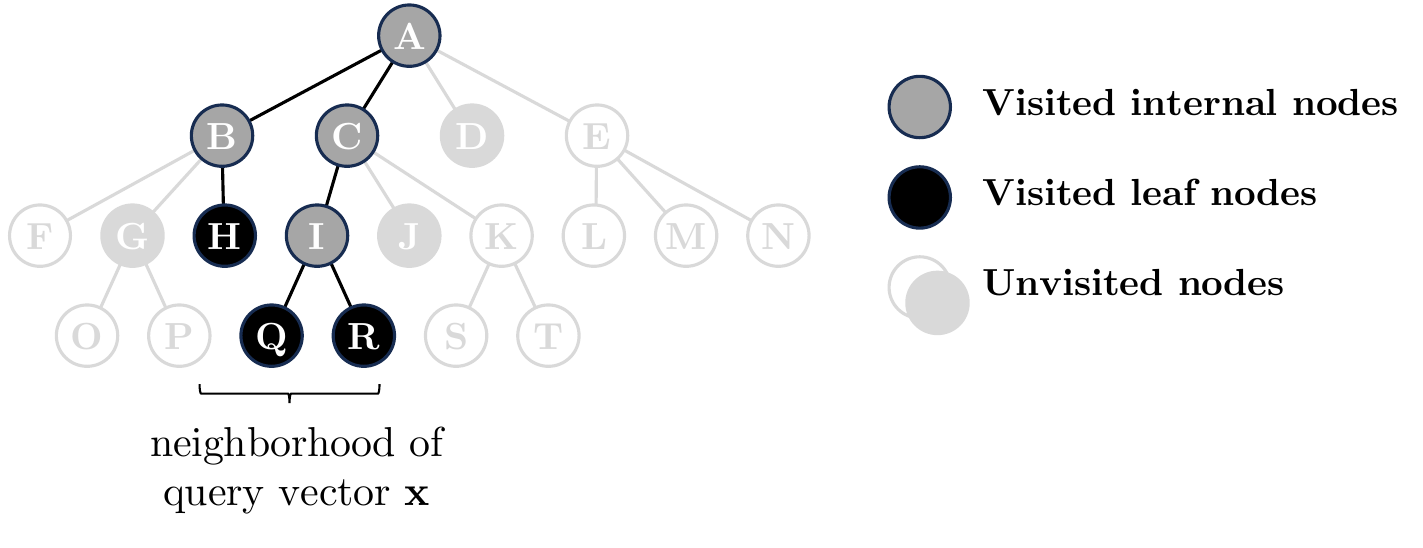}
  \vspace{-5mm}
  \caption{$k$-ANN similarity search on \sys.}
  \vspace{-5mm}
  \label{fig:query}
\end{figure}

\sys conducts a k-ANN search in two stages: (1) Traverse the tenant clustering tree $TCT(t)$ of the querying tenant $t$ to identify the clusters in $IVF(t)$ that are closest to the query vector $x$. (2) Examine these candidate clusters in order of increasing distance to $x$, and scan the shortlists of $t$ within these clusters until a sufficient number of candidate vectors have been found.

When traversing the index tree $I$ starting from the root node, we rely on the Bloom filters and shortlists associated with each node to determine whether we have reached the boundary of $TCT(t)$. If $BF(n)$ indicates that the node does not contain any vectors accessible to the querying tenant (case 1), we determine that we have stepped outside of $TCT(t)$ and backtrack immediately. If the node contains a shortlist of the querying tenant (case 2), then this node should be a leaf of $TCT(t)$ and will be put into the queue of candidate clusters for scanning in the second stage. Otherwise, we must be visiting an internal node of $TCT(t)$ (case 3), so we just expand the search frontier with its children. Note that while Bloom filters are susceptible to occasional false positives, this does not compromise the correctness of the search, but only has a minor effect on search efficiency. In instances where a false positive leads the search algorithm to a subtree outside of $TCT(t)$, it is highly probable that the Bloom filters of the child nodes will correctly return False, thereby halting further unnecessary traversal.

We have already narrowed the search scope from the entire clustering tree $I$ to the nodes within $TCT(t)$ by using Bloom filters and shortlists. However, it is still costly to traverse the $TCT(t)$ exhaustively to identify all $IVF(t)$ clusters, especially when the search algorithm only needs to scan a few of them to achieve high recall. To address this issue, we leverage the tree structure of $TCT(t)$ to further reduce the traversal footprint with a best-first search, prioritizing nodes with centroids closer to the query vector. In this way, the search algorithm quickly locates the nearest cluster to the query vector, and then progressively move away from it to explore other neighboring clusters. The search terminates as soon as a sufficient number of candidate vectors are discovered, often after traversing only a fraction of the $TCT(t)$. Fig.~\ref{fig:query} illustrates this, where the colored nodes represent the querying tenant's TCT, with only the non-grayed nodes being traversed.

We introduce two hyper-parameters, $\gamma_1$ and $\gamma_2$, to control the trade-off between search quality and speed.\footnote{Note that having hyper-parameters is a standard practice in other vector indexes as well in order to control this search quality and search speed trade-off. Similar to other vector indexes, we use grid search on subsampled data to determine the hyper-parameters in \sys.} The imprecision in the search process arises primarily from two factors: (1) k-means clustering intrinsically incurs quantization error, as it represents each vector by the centroid of its assigned cluster; (2) best-first search is not exhaustive for candidate clusters may not return exactly the closest clusters. To mitigate the first source of error, the search algorithm inspects $\gamma_1 \cdot k$ candidate vectors during the shortlist scanning phase before finalizing the top-$k$ results. To address the second, we expand the search radius during the best-first search phase by allowing tree traversal to continue until the cumulative number of vectors within the shortlists found has reached $\gamma_1 \cdot \gamma_2 \cdot k$. While increasing $\gamma_1$ could also compensate for the second source of error, this approach tends to be much more costly because it involves traversing a greater number of shortlists.

\subsection{Index Training}
Similar to the IVF index, \sys requires an initial training phase to construct the GCT, following the same training procedure as the existing literature on hierarchical $k$-means clustering trees~\cite{lamrous2006divisive,muja2009fast}. In this process, we assume a collection of vectors is provided as the training set, which should represent the overall distribution of vectors. After the training phase, the structure of GCT remains fixed as vectors are inserted to or deleted from the index. The per-tenant TCTs are constructed incrementally as tenants are granted/revoked accesses to vectors, as we will describe below. 

Here, fixing the structure of the GCT makes \sys unable to adapt to the drift in the overall distribution of vectors. This is not a fundamental limitation in the design of \sys, but a design choice for the sake of simplicity, assuming the vector distribution will remain relatively stable or re-training will be performed in a batched fashion periodically. These is a rich body of literature on incremental $k$-means~\cite{macqueen1967some}, and these techniques could also be applied to \sys. For example, our evaluation baseline, faiss~\cite{johnson2019faiss} (one of the most popular IVF implementations), also has a fixed IVF tree structure and requires retraining to change the tree structure.

\subsection{Vector Insertion and Access Granting}

\begin{figure}[t]
  \centering
  \includegraphics[width=\columnwidth]{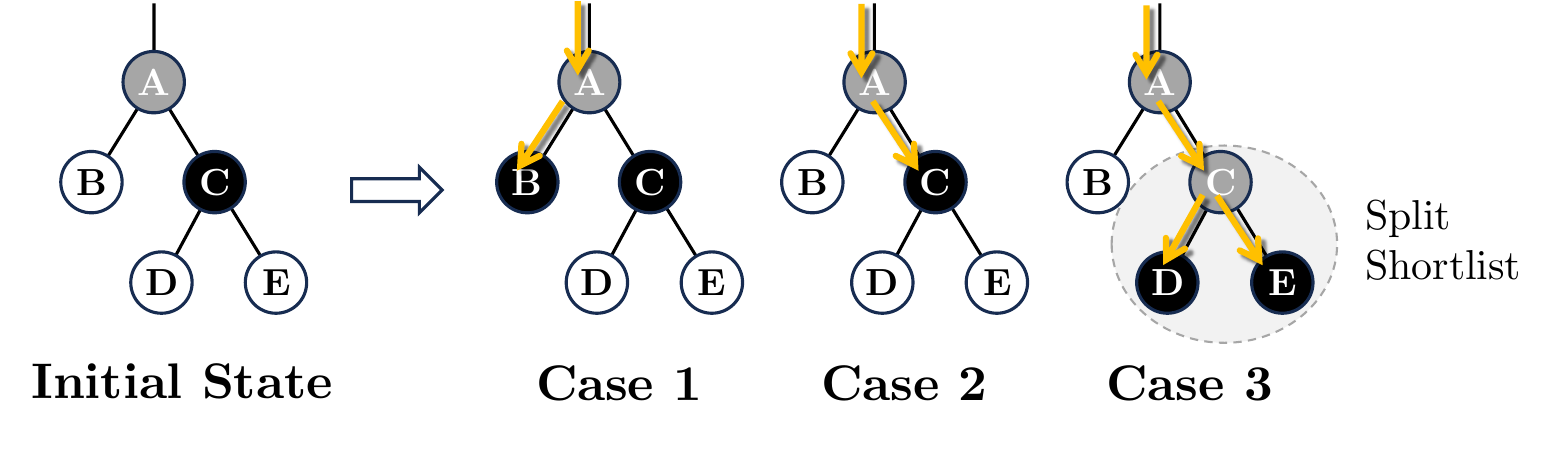}
  \vspace{-5mm}
  \caption{Vector Insertion.}
  \vspace{-5mm}
  \label{fig:insert}
\end{figure}

In \sys, inserting a vector $v$ involves both storing the vector data in the index and granting access to the tenant $t$. The algorithm starts from the root node and greedily traverses toward the nearest leaf node in GCT, where the vector is stored. We refer to this leaf node as the assigned leaf of $v$ and denote it as $leaf(v)$. Because the leaf nodes of GCT represents the most fine-grained partition of the vector space, any cluster in TCTs containing $v$ must be either $leaf(v)$ or its ancestor. Granting access to $t$ requires updating the structure of $TCT(t)$, which is determined by the auxiliary data structured associated with nodes in GCT, including the Bloom filters and shortlists for $t$.

\parab{Updating Shortlists}
During the greedy traversal toward $leaf(v)$, the algorithm determines at each node if it has reached the boundary of $TCT(t)$. This boundary is identified as either an external node where the Bloom filter does not contain $t$, or a leaf node with a shortlist for $t$. If it is the former, a new shortlist is created with $v$ as the only item; Otherwise, $v$ is appended to the existing shortlist.

If a shortlist at node $n$ reaches its maximum capacity during an access granting operation, $TCT(t)$ is expanded by splitting the shortlist among the children of $n$. This step conceptually represents a subdivision of the corresponding cluster in $IVF(t)$ to ensure balanced vector space partitioning. Specifically, each vector in the original shortlist is assigned to the nearest child node, and new shortlists for $t$ are allocated for the child nodes responsible for any vectors. After this step, node $n$ turns into an internal node, and some of its child nodes become new leaf nodes of $TCT(t)$. There are two edge cases we must consider: (1) If node $n$ is a leaf node of GCT and has no child nodes, shortlist splitting is not feasible. Hence, we do not impose a size limit on the shortlists at GCT's leaf nodes. (2) All vectors within the original shortlist might be assigned to the same child node, which means it is simply moved to the child node and remains overfull. In this case, we recursively split it until either it is successfully divided or a leaf node of GCT is reached.

\parab{Updating Bloom Filters}
Recall that the Bloom filter at node $n$ represents the set of tenants who have shortlists in the sub-tree rooted at $n$. Since shortlists for a specific tenant only exist at the leaf nodes of its corresponding TCT, the output of Bloom filter equivalently determines whether node $n$ belongs to TCT. To uphold such property, when a new shortlist for $t$ is created at a node, turning it into a leaf node of $TCT(t)$, the Bloom filters at that node must be updated to include $t$. By induction, this step ensures that the Bloom filter at every node of $TCT(t)$ includes $t$.

Fig.~\ref{fig:insert} illustrates an example of granting access to vector $v$ for tenant $t$. Fig.~\ref{fig:insert}a shows the initial state of $TCT(t)$ and GCT. In case 1, the greedy traversal process reaches node $B$, an external node of $TCT(t)$. Thus, the algorithm creates a new shortlist for $t$ and updates its Bloom filter. In case 2, a leaf node $C$ with a non-full shortlist is reached, so the algorithm simply appends $v$ to the shortlist. In case 3, the shortlist at $C$ is at capacity, so the algorithm performs shortlist splitting after appending $v$ to it. This splitting step distributes the vectors in the shortlist among two children $D$ and $E$, which become new leaf nodes of $TCT(t)$.

\subsection{Vector Deletion and Access Revocation}

\begin{figure}[t]
  \centering
  \includegraphics[width=\columnwidth]{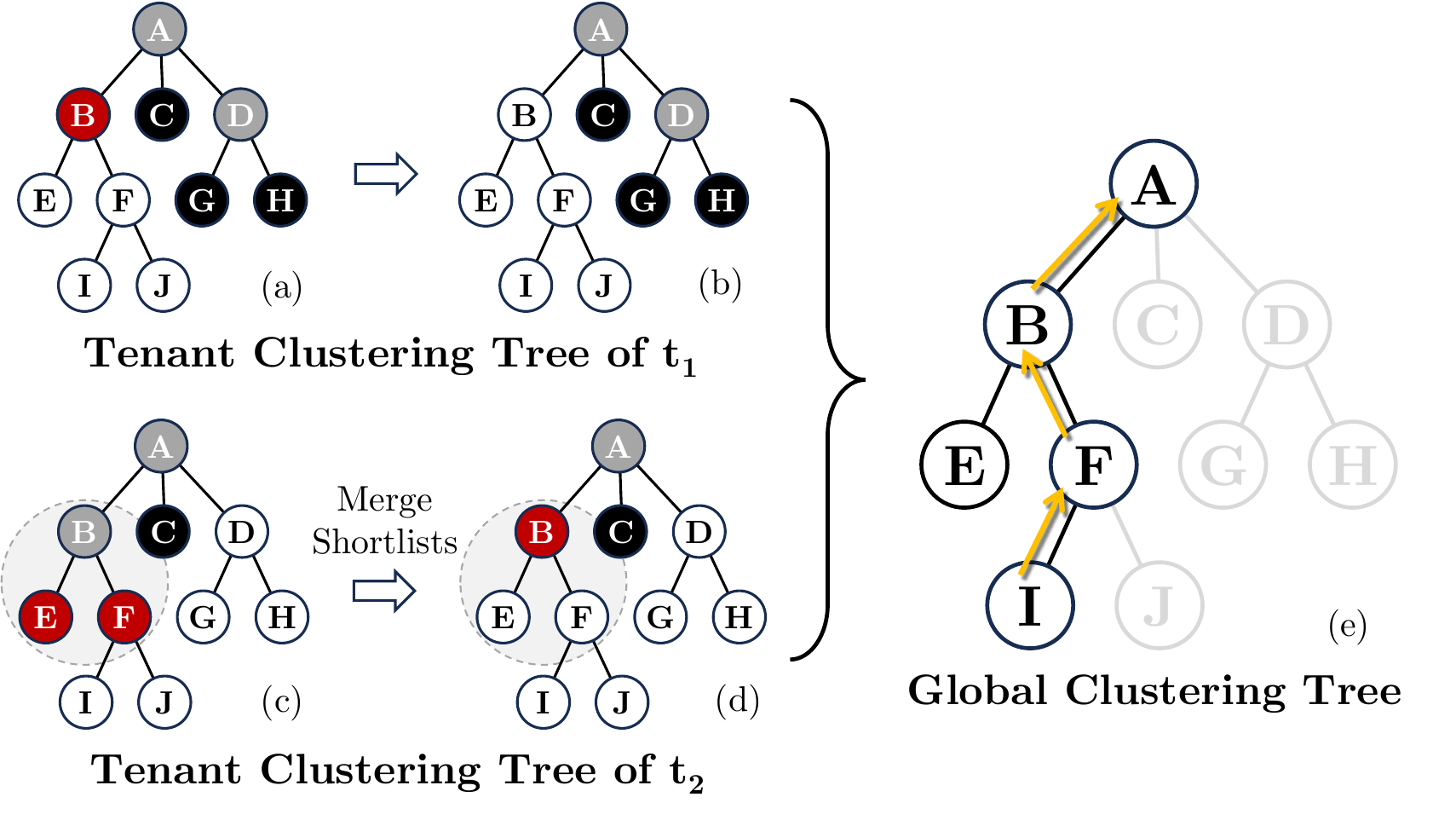}
  \vspace{-5mm}
  \caption{Vector Deletion.}
  \vspace{-5mm}
  \label{fig:delete}
\end{figure}

Deleting a vector in \sys involves revoking access from all tenants on its access list in a batched fashion, and then removing the vector data from the index. The access revocation operation updates the corresponding TCT, which, as discussed in \autoref{sec:overview}, is defined by the Bloom filters and shortlists associated with each node of the GCT. In this section, we first explain how Bloom filters and shortlists are updated during an access revocation operation, and then describe how a vector deletion operation carries out batch access revocation and vector data removal. For a clearer demonstration, we will use the example shown in Fig.~\ref{fig:delete} throughout this section. In this example, we are deleting a vector $v$ assigned to leaf node $I$ of GCT. On the left side of the figure, we illustrate the process of revoking access from two tenants $t_1$ and $t_2$; On the right side (Fig.~\ref{fig:delete}e), we show how we perform batch access revocation.

\parab{Updating Bloom Filters}
The primary goal of the access revocation operation is to keep the corresponding tenant $t$'s TCT up-to-date. This goal can be further broken down into two sub-tasks: (1) Maintaining an correct structure of the TCT, i.e., how $IVF(t)$ partitions the vector space; (2) Updating the content of shortlists at the TCT's leaves, i.e., the set of vectors within each cluster of $IVF(t)$. Since every TCT shares tree structure with GCT and grows from the same root, maintaining the structure of a TCT is equivalent to maintaining its boundary. Recall that the boundary of a TCT is determined by two scenarios: (1) The Bloom filter returns false for $t$, suggesting this node is outside of the TCT; (2) The node contains a shortlist for tenant $t$, indicating this node is a leaf node of the TCT. Ensuring correctness in these two scenarios require correctly updating the Bloom filters and shortlists, respectively.

Updating Bloom filters is only necessary when the access revocation operation deletes the last vector from a shortlist, leading to the shortlist's removal. In this case, the node containing the shortlist, denoted as $n$, turns from a leave node into an external node of the TCT, and tenant $t$ should be removed from the associated Bloom filter by definition. Eliminating an element from Bloom filters generally involves re-computation from scratch, which, in our situation, requires locating all shortlists in descendant nodes. However, the hierarchical structure of the index allows us to efficiently re-compute the Bloom filter by merging the Bloom filters of child nodes and adding the remaining shortlists in $n$. Additionally, we also need to recursively update the Bloom filters of the ancestors of $n$, which can also be implemented through merging Bloom filters. The recursive update process ends when an ancestor node's Bloom filter remains unchanged after re-computation.

For example, Fig.~\ref{fig:delete}a and Fig.~\ref{fig:delete}b illustrates an access revocation operation that involves updating Bloom filters. The algorithm starts by traversing from $I$ to the root node to find a shortlist for tenant $t_1$. It then locates the shortlist at node $B$ and removes $v$ from it. Following the removal, it finds out the shortlist is empty, so it removes the shortlist from $B$ and re-computes the Bloom filter of $B$ as $BF(E) \cup BF(F) \cup SL(B)$. It continue by re-computing the Bloom filter of the parent node $A$ as $BF(B) \cup BF(C) \cup BF(D) \cup SL(A)$. The Bloom filter of node $A$ will remain unchanged after re-computation because its other sub-trees contain shortlists for $t_1$, so the recursive update process will end.

To further reduce the overhead of updating Bloom filters, the Bloom filter at a node can be re-computed only once in several consecutive updates. Each re-computation refreshes the Bloom filter to reflect all delayed updates, effectively batching them. The impact of batched updates on other operations is similar to the false positives in Bloom filters as discussed in prior sections, i.e., it only slightly affects performance while preserving correctness.

\parab{Updating Shortlists}
After identifying the target shortlist for tenant $t$, we typically only need to remove vector $v$ from it. However, sometimes additional work is required to maintain the index structure. Recall that a shortlist will be divided if its size exceeds a certain threshold after an insertion or access granting operation. Conversely, when the total size of shortlists for tenant $t$ within a sub-tree falls below the threshold after a deletion or access revocation operation, these shortlists should be merged back into one. At first sight, checking this condition appears complicated: (1) Computing the total size of shortlists in a sub-tree requires a tree traversal to locate all the shortlists; (2) There could be several nested sub-trees that meet the condition, and the shortlists should be merged up to the root of the largest one. 

To address the first issue, we rely on the hierarchical structure of the index. Instead of traversing the entire sub-tree, we can determine the total size of shortlists just from the direct children. Suppose we are trying to merge the shortlists of $t$ in a sub-tree rooted at node $n$. In this case, we need to calculate the total number of vectors accessible to $t$ in node $n$, which is equivalent to the total size of $t$'s shortlists in its descendants. Given that the child nodes of $n$ represents a further partition of the cluster represented by $n$, we can calculate the total number of vectors in $n$ by adding up those in the children nodes. A node's type inherently indicates the number of $t$’s vectors it contains: external nodes contain no vectors, leaf nodes contain fewer than threshold vectors (otherwise it would be split and become an internal node), and internal nodes contain more than threshold vectors (otherwise, merging shortlists would have turned it into a leaf node). Hence, an eligible node $n$ should meet two criteria: (1) None of child nodes is internal; (2) The total size of shortlists in its direct children is below the threshold.

We address the second issue by merging shortlists recursively. Assume node $n$ contains the updated shortlist. We begin by checking the sub-tree rooted at the parent node of $n$, which we denote as $p$. If it is deemed eligible, we merge the updated shortlist along with the shortlists in $n$'s sibling nodes into the parent node. Following this, we proceed to examine the parent node of $p$, and if it meets the criteria, we execute the merging process again. This procedure is repeated until the condition is no longer met. 

As shown in Fig.~\ref{fig:delete}c and Fig.~\ref{fig:delete}d, shrinking the TCT of tenant $t_2$ during access revocation requires shortlist merging. After $v$ is removed from $SL(F, t)$, the shortlists in node E and F are merged into node B. Since the total size of the merged shortlist and $SL(C, t)$ is larger than the threshold, the recursive merging process stops. 

\section{Implementation}
We develop \sys in approximately 1000 lines of C++ on top of the \texttt{faiss} library~\cite{johnson2019faiss}. We rely on OpenMP~\cite{chandra2001parallel} to implement multi-threaded search and the C++ Bloom Filter Library~\cite{bloomfilter} for Bloom filter implementation. 

\subsection{API Design}
\sys provides the same set of APIs as today's vector indexes besides a few changes to consider multi-tenancy.
During setup, the \texttt{train\_index (vector[], params)} API takes a representative set of vectors to construct the clustering tree. The construction parameters constrain the shape of the clustering tree, such as tree depth and branching factor. 

Once the index is trained, the application can insert vectors into the index using the \texttt{insert\_vector(vector, label, tenant)} function. The insertion operation associates each vector in the index with a unique label, which acts as a reference for stored vectors, and assigns ownership to the specified tenant. Conversely, the \texttt{delete\_vector(label)} removes a vector from the index based on its label and also implicitly revokes access for all tenants in the access list. \sys further provide \texttt{get\_vector(label)} to retrieve the vector data using the associated label.

For permission management, \sys provides \texttt{grant\_access (label, tenant)} and \texttt{revoke\_access(label, tenant)} to grant and revoke access of a vector to/from a tenant. These two functions modify the access list associated with the vector, which initially contains only the owner. The application can check whether a tenant has access or ownership for a vector using \texttt{has\_access(label, tenant)} and \texttt{has\_ownership(label,tenant)}.

To perform k-NN queries for a tenant, the application can call the \texttt{knn\_search(vector, k, tenant, params)} function. Here, \texttt{params} represents the hyper-parameters that controls the search radius, including $\gamma_1$ and $\gamma_2$, as discussed in \autoref{sec:algorithm:search}.

\subsection{Parallel Vector Search on Multiple Cores}

We support multi-threaded search in \sys in two modes, namely inter-query parallelism and intra-query parallelism. In the inter-query parallelism mode, a batch of search queries is dynamically scheduled among a pool of worker threads in a first-come-first-serve manner. This mode can parallelize the entire query execution process, but cannot reduce query latency. On the other hand, the intra-query parallelism mode focuses on reducing the latency of a single query. In this mode, worker threads work in parallel to scan different candidate clusters. To further increase parallelism and improve load balance, we divide vectors in each cluster into fixed-size chunks and distribute them among workers. A chunk size of 16 works best in most cases. Due to the overhead of multi-threading and implementation complexity, we do not parallelize the first stage of the search process, where the search algorithm traverses the clustering tree to identify candidate clusters.

\section{Evaluation}

\begin{table}[t]
\centering
\caption{Characteristics of Evaluation Datasets}
\vspace{-3mm}
\label{tab:dataset_characteristics}
\begin{tabular}{lcc}
\toprule
                   & \textbf{YFCC100M}  & \textbf{arXiv} \\
\midrule
\textbf{Number of Vectors}            & 1M      & 2M        \\
\textbf{Vector Dimensions}            & 192     & 384       \\
\textbf{Number of Tenants}            & 1000    & 100       \\
\textbf{Avg Sharing Degree}           & 13.37   & 9.93       \\
\bottomrule
\end{tabular}
\vspace{-3mm}
\end{table}

In this section, we first evaluate the overall system performance for key database operations like search, insertion, and deletion. We then examine how \sys scales to different query selectivities and numbers of tenants. Finally, we analyze the contribution of individual system components.

\subsection{Experimental Setup}
\label{sec:exp_setup}

We use a server with an Intel Xeon Gold 5215 @2.5GHz processor and 256GB memory, running Linux Ubuntu 20.04 LTS.

\parab{Baselines.}
We compare the performance of our index against two of the most widely used vector indexes: the partition-based IVF-Flat index~\cite{johnson2019faiss} and the graph-based HNSW index~\cite{malkov2018hnsw}. 

We implemented \sys and the IVF baselines based on \texttt{faiss v1.7.4}~\cite{johnson2019faiss}, and HNSW baselines based on \texttt{hnswlib v0.7.0}~\cite{hnswlib}. Both \texttt{faiss} and \texttt{hnswlib} support single-stage metadata filtering, so we only need to provide a permission checking predicate to these two libraries.
For per-tenant baselines, we used the original IVF-Flat and HSNW indexes and added logic that route queries to the index of the querying tenant. For all indexes, we store vector in their original representation without scalar or product quantization.

In the following sections, we refer to these baselines as PT-IVF/HNSW and MF-IVF/HNSW. PT and MF are abbreviations for per-tenant indexes and metadata filtering, respectively.

\parab{Workloads.} Since existing vector search benchmarks do not have information about tenant-specific access permissions, we generate vectors and access metadata from some well-known datasets.

The \textbf{YFCC100M} dataset comprises 1M images randomly chosen from the original YFCC100M dataset~\cite{thomee2016yfcc100m}. This dataset contains a hybrid of vector and scalar data, with images embedded using the CLIP model~\cite{radford2021learning} and annotated with tags that denote entities, locations, etc. To create a access matrix that reflects the natural characteristics in each tenant's vector distribution, we grant each tenant the access to all images associated with a specific tag. For instance, a tenant specializing in landscape photography only has access to photos tagged as ``landscape''. We consider only the 1000 tags that appear most frequently in the dataset so that both \sys and the baseline systems' indexes can fit into the memory of our server.

The \textbf{arXiv} dataset~\cite{clement2019usearxiv} contains metadata of 2.3M arXiv papers. This includes information such as titles, authors, abstracts, and paper categories (e.g., \verb|cs.DB| denotes the field of database systems). We generate embedding vectors by transforming the abstract of each paper using the \verb|all-MiniLM-L6-v2|~\cite{huggingface_allMiniLM-L6-v2} model. The access matrix is created similarly to the YFCC100M dataset, where we assume each tenant only access papers from a few categories. This is typical in a realistic scenario, because most researchers typically only search papers in the fields most related to their interests. We use only the 100 largest categories in the dataset as tenants.

\subsection{System Evaluation}
\label{sec:sys_eval}

We first evaluate the performance of standard manipulation and search queries of \sys compared to the previously described baselines. Unless otherwise specified, in each experiment below, we select index configurations that achieve the best search performance while maintaining a recall of at least 95\% through grid search.

We conduct the benchmark experiment as follows. First, we randomly split each dataset into a training set and a test set of size 10000. For indexes that require an initial training process (including \sys and the IVF-based baselines), we use all vectors in the training set to train the index. Once the index has been initiated, we sequentially insert vectors in the training set into the index. Each insertion operation involves adding a new vector to the index and granting access permissions to every tenant in its access list. To evaluate search performance, we execute search queries for every $\langle vector, tenant \rangle$ pair in the test set. Finally, vectors are gradually removed from the index. Each deletion operation involves revoking access for all tenants in the access list and eventually erasing the vector data. All operations are executed sequentially without batching, with the exception of multi-threaded search.

\begin{figure}[t]
  \centering
  \includegraphics[width=\linewidth]{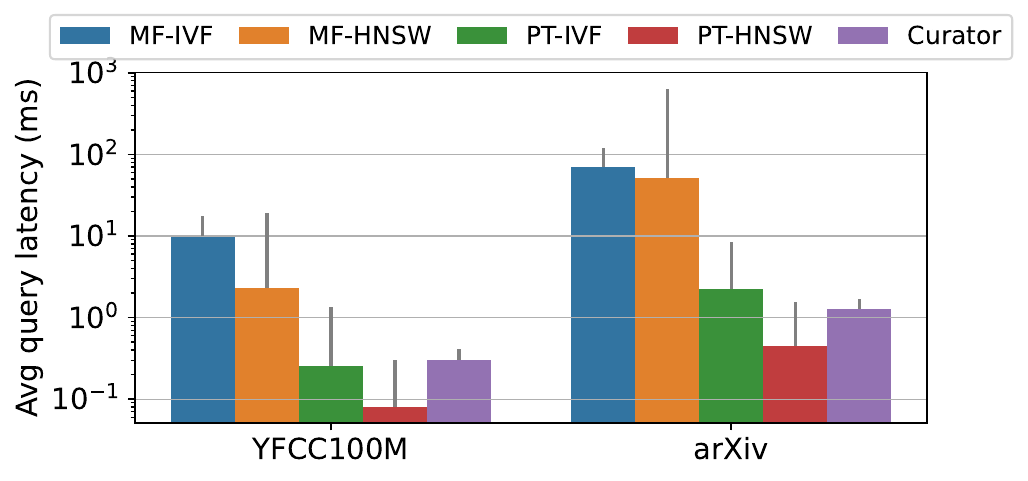}
  \vspace{-5mm}
  \caption{Performance comparison of k-NN queries. \texttt{MF} denotes metadata filtering and \texttt{PT} denotes per-tenant indexing. The top of the error bar represents the P99 latency.}
  \vspace{-4mm}
  \label{fig:query_performance}
\end{figure}

\begin{figure}[t]
  \centering
  \includegraphics[width=\linewidth]{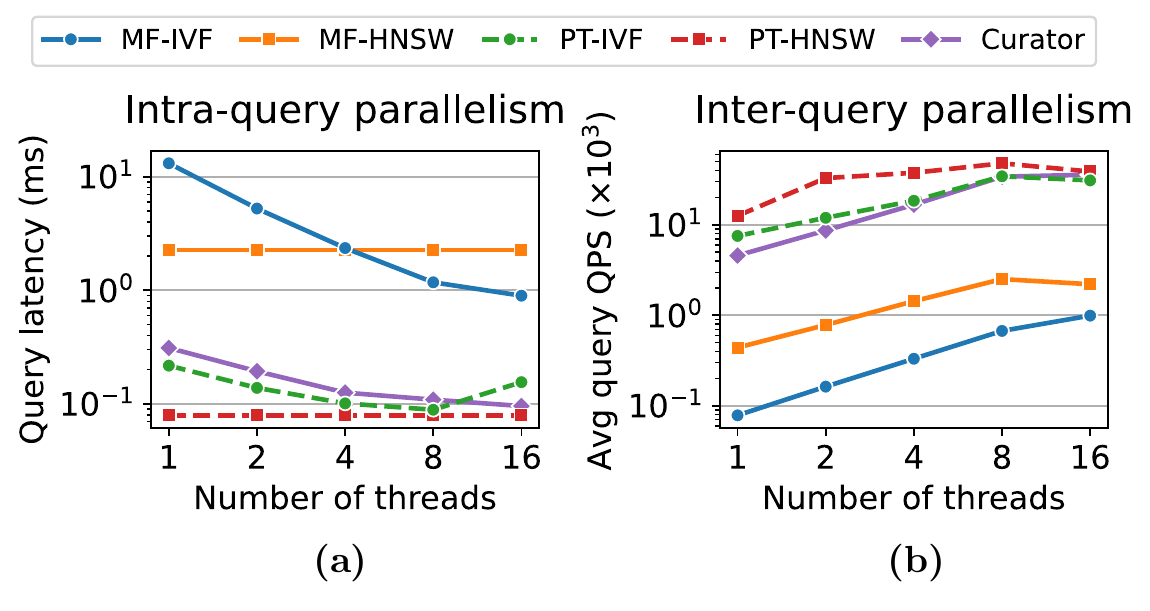}
  \vspace{-5mm}
  \caption{Multi-threaded query performance on YFCC100M. Since HNSW baselines does not support intra-query parallelism, single-threaded latencies are plotted for reference.}
  \vspace{-3mm}
  \label{fig:mt_query_performance}
\end{figure}

\parab{Query Performance} As shown in Fig.~\ref{fig:query_performance}, \sys significantly outperforms the metadata filtering baselines in query latency on both datasets. For the YFCC100M dataset, \sys executes queries 32.9$\times$ faster than MF-IVF and 7.7$\times$ faster than MF-HNSW. The performance advantage generalizes to the arXiv dataset, where \sys achieves query latencies that are 54.6$\times$ and 40.3$\times$ lower than MF-IVF and MF-HNSW, respectively. When compared to per-tenant index baselines, \sys shows comparable performance to the PT-IVF index, but is a factor of 3.3 to 4.7 slower compared to the PT-HNSW baseline. We attribute this performance gap to the inherent trade-off between memory consumption and query speed --- while the PT-HNSW baseline prioritizes fast querying at the expense of much higher memory usage (as shown below), \sys achieves a better balance between resource efficiency and performance, catering to the real-world deployment scenarios where memory overhead is a critical consideration. P99 latency has the same overall trends as average latency when comparing \sys to baselines.

In latency-critical scenarios, intra-query parallelism can be applied to bridge the latency gap, as shown in Fig.~\ref{fig:mt_query_performance}a. 
Since intra-query parallelism is challenging for graph-based vector indexes and not supported by open-sourced HNSW implementations, we only plot single-thread search latency for HNSW baselines as a reference. For the \sys and IVF baselines, we distribute the task of scanning IVF clusters among multiple threads, which reduces the query latency significantly.
By using 16 threads, the performance gap between \sys and PT-HNSW can be narrowed to just 20\% (16us).

We next evaluate the query throughput via inter-query parallelism. 
For all indexes, we input queries from each tenant in a single batch and dynamically schedule these queries among a pool of worker threads.
Fig.~\ref{fig:mt_query_performance}b shows the search throughput of \sys and IVF baselines scales better with the number of threads compared to the HNSW variants, with the performance gap between \sys and PT-HNSW quickly diminishing for $>8$ threads.

\begin{figure}[t]
  \centering
  \includegraphics[width=\linewidth]{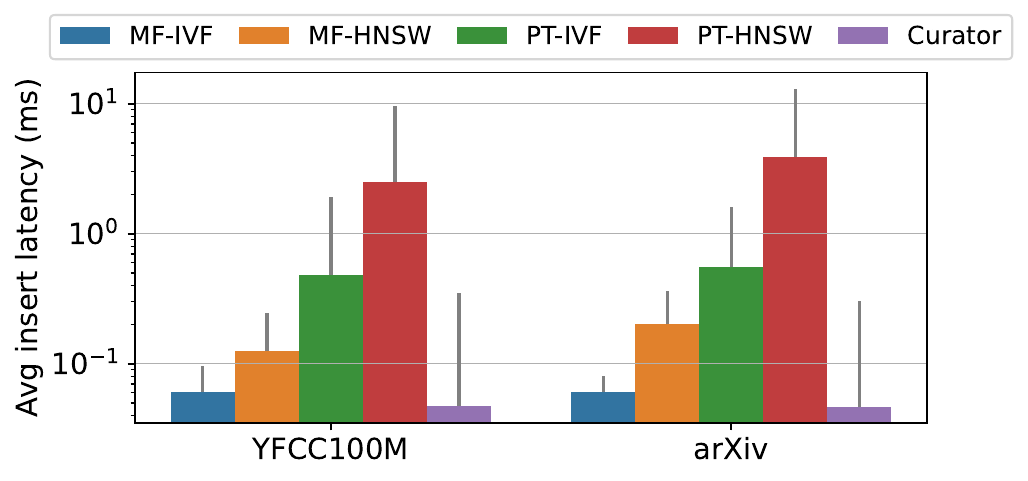}
  \vspace{-5mm}
  \caption{Performance comparison of insertion operations. Each insertion operation involves adding a new vector to the index and granting accesses to all tenants in the access list. The top of the error bar represents the P99 latency.}
  \vspace{-3mm}
  \label{fig:insert_performance}
\end{figure}
 
\begin{figure}[t]
  \centering
  \includegraphics[width=\linewidth]{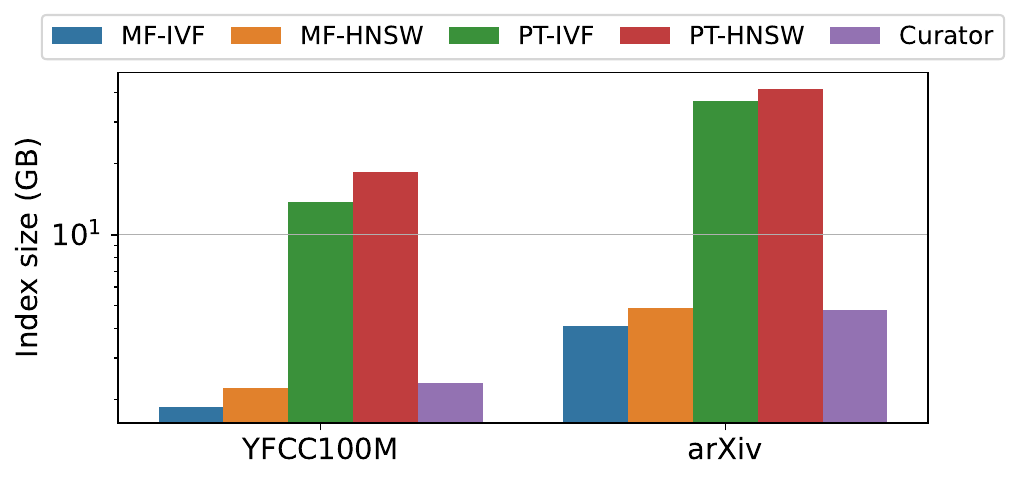}
  \vspace{-5mm}
  \caption{Index size comparison}
  \vspace{-3mm}
  \label{fig:index_size}
\end{figure}

\parab{Memory Consumption.}
Fig.~\ref{fig:index_size} shows how much memory is needed for different indexes to store the YFCC100M and arXiv datasets. The memory usage of \sys is 7.9-8.7$\times$ lower than separate HNSW and 5.9-7.6$\times$ lower than separate IVF-Flat. This reduction is mostly due to eliminating duplicate vector data in per-tenant indexes. In addition, separate HNSW also suffers from the overhead of redundant edges, where each occurrence of a shared vector in different indexes maintains its own unique set of edges to adjacent nodes. Compared to metadata filtering baselines, \sys incurs minimal memory overhead, by eliminating vector duplication and introducing minimal auxiliary data structures to speed up queries.

\parab{Insertion/Deletion Performance.}
Fig.~\ref{fig:insert_performance} compares the insertion latency of \sys to the baselines. Each insertion operation involves introducing a new vector to the index and granting accesses to tenants. For both metadata filtering and per-tenant index strategies, HNSW achieves much lower performance than IVF due to the overhead of incremental graph construction. Besides, per-tenant indexes are much slower because each insertion operation needs to update multiple indexes. The insertion speed of \sys is even higher than the MF-IVF baseline, as the hierarchical clustering tree allows us to minimize the number of distance computations required to find the cluster that the new vector belongs to.

\begin{figure}[t]
  \centering
  \includegraphics[width=\linewidth]{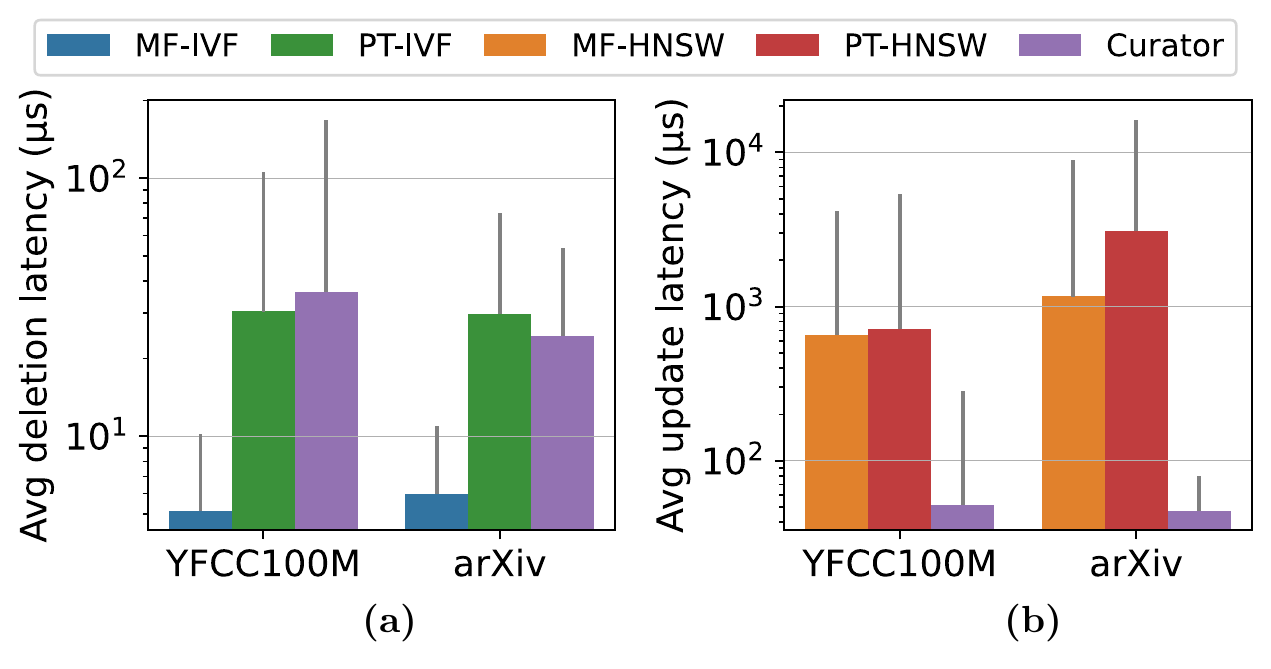}
  \vspace{-8mm}
  \caption{Performance comparison of deletion operations. We compare the update performance against HNSW baselines, because in HNSW indexes, data associated with a deleted vector remains in the index until it is overwritten by a newly inserted vector occupying the same slot. The top of the error bar represents the P99 latency.}
  \vspace{-5mm}
  \label{fig:delete_performance}
\end{figure}

Unlike other indexes that remove vector information instantly during a deletion operation, the HNSW index typically defers this process. Rather than immediately erasing the vector data and its connected edges, the index marks the vector as deleted and excludes it from future searches. When a new vector is introduced to the index, the slot of the old vector may be reused. When this occurs, the outdated vector data will be overwritten by the new one, and the edges in the neighborhood of the old vector will be updated to reflect the removal.
To ensure fairness, we compare the performance of vector deletion with IVF baselines, and the performance of vector update with HNSW baselines. For all indexes, the vector update operation is implemented as a deletion followed by an insertion. As shown in Fig.~\ref{fig:delete_performance}a, all three indexes achieve high deletion performance since distance computation is not required in this process. Among them, the deletion latency of \sys is close to PT-IVF but higher than MF-IVF, as both \sys and PT-IVF need to update all tenants in the access list. Fig.~\ref{fig:delete_performance}b shows that the overhead of updating vectors in HNSW baselines is much higher than \sys. This is because HNSW indexes need to repair the graph structure in the neighborhood of the deleted vector, which involves many distance computations. Note that P99 latency for insertion and deletion is significantly higher than the average across \sys and other baselines. This is because some updates requires significant modifications to the index at runtime (e.g., merging and spliting of shortlists in \sys).

\subsection{Scalability}

\begin{figure}[t]
  \centering
  \includegraphics[width=\linewidth]{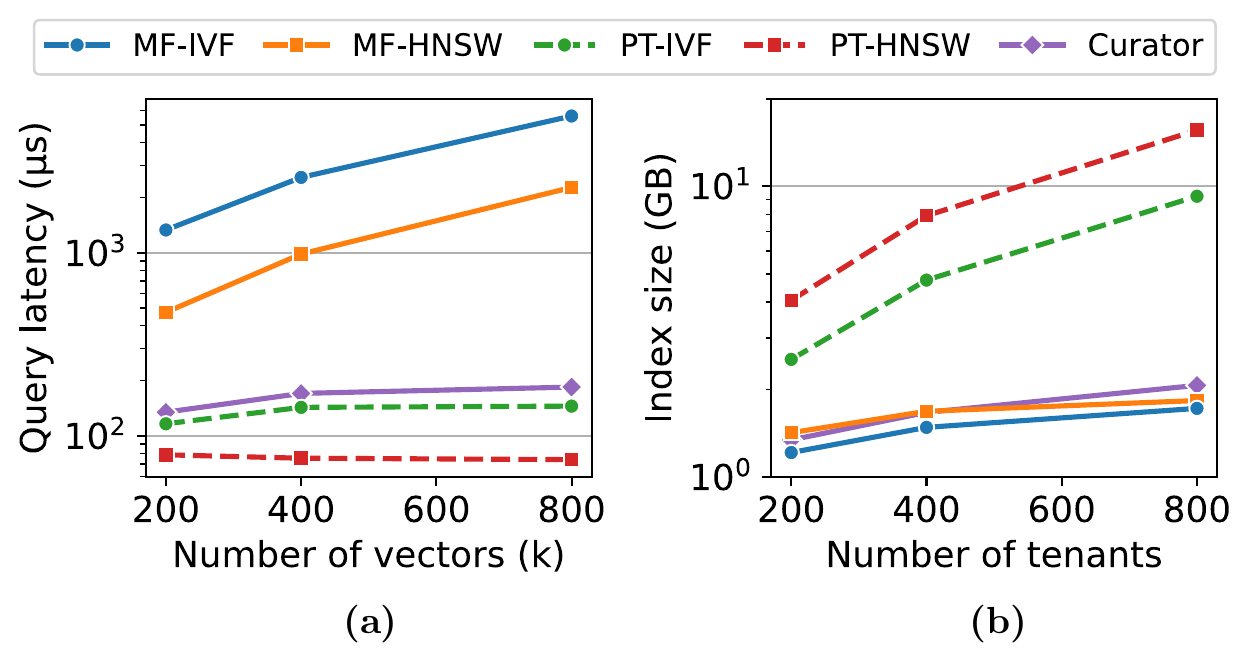}
  \vspace{-8mm}
  \caption{\sys demonstrates high scalability across various dataset sizes and numbers of tenants. In our experimental setup, a larger dataset size leads to lower query selectivity, and more tenants correspond to a higher sharing degree.}
  \vspace{-5mm}
  \label{fig:ablation_scalability}
\end{figure}

In this section, we create synthetic datasets to have better control over their characteristics. These datasets borrow vectors from the YFCC100M dataset, but randomly assign the vectors to tenants. We use the same configuration for all indexes, as in \autoref{sec:sys_eval}. 

\parab{Query Selectivity.}
For the sake of simplicity, we refer to the proportion of vectors a certain tenant can access in all vectors as "query selectivity". This is a slight abuse of terminology, because multi-tenant vector search is not necessarily implemented with metadata filtering. A well-designed multi-tenant index should maintain stable search performance for each tenant, regardless of how much the total number of vectors increases. In other words, the search performance experienced by one tenant should not be significantly impacted by the data growth of another. This can be seen as a limited form of performance isolation. 

To evaluate how \sys performs at different query selectivity levels, we keep the number of tenants and the number of vectors each tenant can access constant, while increasing the total number of vectors. Fig.~\ref{fig:ablation_scalability}a shows that both \sys and the per-tenant indexes maintain roughly the same query latency across all query selectivities. In contrast, indexes based on metadata filtering experience a significant drop in search performance when the selectivity is low, due to the overhead of examining irrelevant vectors.

\parab{Number of Tenants.}
As a multi-tenant vector database has more tenants, the per-tenant memory overhead limits the maximum number of tenants that can be hosted on the same machine. To assess how the index size of \sys scales with the number of tenants, we generate a series of synthetic datasets with an increasing number of tenants, while keeping the total number of vectors and the number of vectors each tenant can access constant. This is equivalent to increasing the sharing degree of each vector.

As shown in Fig.~\ref{fig:ablation_scalability}b, the memory usage of \sys and the indexes based on metadata filtering increases slowly in relation to the number of tenants, as they only maintain minimal information for each tenant. In contrast, the per-tenant indexes quickly bloats the total memory consumption due to the overhead of creating a standalone vector index for each new tenant.

\subsection{Ablation Study}
\label{sec:ablation}

\begin{figure}[t]
  \centering
  \includegraphics[width=\linewidth]{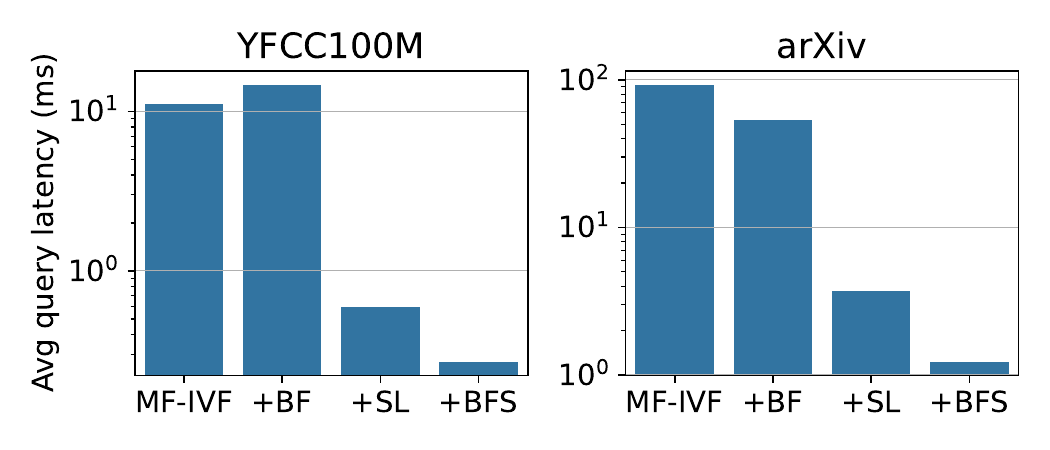}
  \vspace{-5mm}
  \caption{How each component contributes to \sys's search performance. The terms \texttt{+BF}, \texttt{+SL} and \texttt{+BFS} refer to MF-IVF augmented with Bloom filters, \sys without the best-first search, and \sys, respectively.}
  \vspace{-3mm}
  \label{fig:ablation}
\end{figure}

\parab{Drilling Down into Components.}
We now evaluate each component's contribution to \sys's search performance. Starting with the MF-IVF baseline, we gradually extend it until it becomes equivalent to \sys:

\textit{Flat IVF-BF (+BF)}. Flat IVF-BF associates each cell of the MF-IVF index with a Bloom filter. During a search, it skips the cells that do not contain any accessible vectors using the Bloom filters. The remaining cells are then scanned with metadata filtering in the same way as MF-IVF. 

\textit{Flat IVF-BF w/ shortlists (+SL)}. In addition to Bloom filters, this index also maintains shortlists in each cell to reduce the cost of examining access lists during metadata filtering.

\textit{Curator (+BFS)}. \sys further enhances search performance by using best-first search to approximately determine the nearest clusters. As the search halts immediately when a sufficient number of candidate vectors are found, it avoids the overhead of traversing the entire clustering tree.

Fig.~\ref{fig:ablation} shows the search latency of the above indexes. Both shortlists and best-first search boost performance significantly by avoiding scanning access lists and exhaustive tree traversal. However, Bloom filters alone provide limited contribution to search performance. This is because the number of cells that can be skipped depends on the data distribution of the querying tenant. In the scenarios where it is low, the cost of querying Bloom filters may outweigh the advantages of cell skipping.

\begin{figure}[t]
  \centering
  \includegraphics[width=\linewidth]{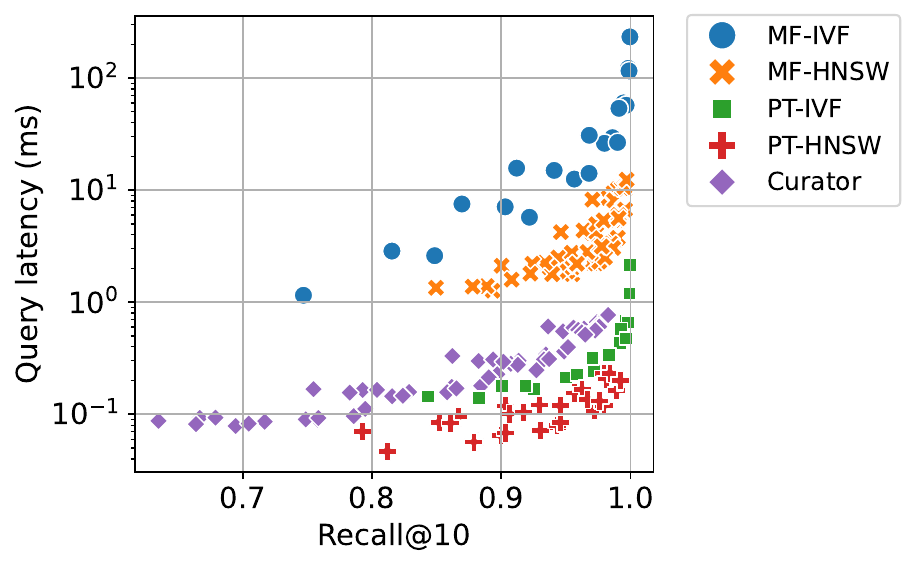}
  \vspace{-5mm}
  \caption{Recall-vs-latency trade-off on YFCC100M dataset. Lower and further to the right is better.}
  \vspace{-5mm}
  \label{fig:param}
\end{figure}

\parab{Recall-vs-Latency Trade-off.}
In this section, we evaluate \sys's search latency against various baseline indexes at different recall levels. The results, as shown in Fig.~\ref{fig:param}, reveal that \sys consistently outperforms metadata filtering baselines across all recalls. When compared to per-tenant index baselines, \sys delivers comparable results to PT-IVF and maintains a small performance gap against PT-HNSW, without incurring the significant memory overhead of per-tenant index baselines.

\section{Related Work}

\parab{Vector indexes.}
As vector similarity search becomes an increasingly important operation in various modern applications, many efficient vector indexes have been proposed, including graph-based indexes~\cite{malkov2018hnsw,fu2019nsg,fu2016efanna,harwood2016fanng,iwasaki2018optimization,jayaram2019diskann} and partition-based indexes, which can be further subdivided into clustering-based~\cite{johnson2019faiss,babenko2014inverted,baranchuk2018revisiting,zhang2014composite}, hashing-based~\cite{gionis1999similarity,andoni2015practical,xu2011complementary,shrivastava2014improved,shrivastava2014asymmetric,neyshabur2015symmetric} and tree-based~\cite{muja2009fast,annoy,muja2014scalable}. Different types of vector indexes represent different trade-offs in terms of search speed, index size and construction time, etc. Another key difference is different index design can require significant considerations when executing on multi-core machines.
Partition-based index (e.g., \sys, IVF) can be easily executed in parallel on multiple cores for intra-query parallelism. This can significantly reduce the query latency (as shown in \autoref{sec:sys_eval}). However, graph-based indexes (e.g., HNSW) are much harder to be parallelized to reduce query latency. Inter-query parallelism (i.e., running different queries on multiple cores) is easy for both types of indexes, however, it only increases query throughput but does not reduce query latency. These traditional index design tries to adapt to data distribution to facility efficient search. Our approach to adapt the index to data distribution is different. These approaches build the index from scratch, while we let each TCT fit a tenant's distribution by choosing different subset of the GCT. Our approach is the key to enable encoding many TCTs on a single GCT with a low memory footprint.

While the primary goal of these indexes is to accelerate vector similarity search by reducing the number of distance computations, typical vector indexes also optimize for other practical objectives. For example, vector indexes may employ compressed vector representations~\cite{jegou2010product} and disk-oriented design~\cite{jayaram2019diskann} to cater for limited memory; some indexes are specialized for write-heavy workloads by supporting real-time updates~\cite{sundaram2013streaming,singh2021freshdiskann}. However, these works focus on single-tenant use cases, where \sys targets at efficient index design for multi-tenant vector databases.

\parab{Multi-tenant databases.}
In recent years, the rise of the Software-as-a-Serivce (SaaS) model has led to growing interest in multi-tenant database systems. 
Research in this field focus on several key themes: the extensibility and flexibility of multi-tenant architectures that allows for tenant-specific modifications~\cite{aulbach2011extensibility,aulbach2008multi,weissman2009design,aulbach2009comparison}; the scalability to an increasing number of tenants with bounded query performance degradation~\cite{hui2009supporting}; robust security and access control mechanisms~\cite{li2006dynamic,sion2005query}; facilitating data sharing among tenants, which is crucial for collaboration and integration~\cite{aulbach2011extensibility}. 

There are overall three approaches to implementing multi-tenant databases~\cite{hui2009supporting}: (1) running independent database instances to serve different tenants, (2) creating private tables for each tenant in a shared instance, and (3) storing tuples of all tenants in a shared table and differentiate them with a special \texttt{TenantID} attribute. The first approach achieves best data isolation and security, but consumes much more memory and computation resources and incurs high maintenance cost; The second approach trades data isolation for lower maintenance cost, but the high resource consumption limits its scalability to a large number of tenants; The last approach achieves highest scalability because its memory footprint is independent of the number of tenants, but the shared index could be inefficient for individual tenants with diverse attributes and access patterns. 

Current vector databases~\cite{pinecone,wang2021milvus,weaviate,qdrant} generally provide the same set of multi-tenancy strategies and leave it to the users to navigate the trade-offs. In our discussion, we categorize both of the first two approaches into the per-tenant indexing strategy and refer to the last one as the metadata filtering strategy. To the best of our knowledge, no vector database has attempted to improve the trade-off by redesigning the vector index.

\section{Conclusion}
This paper introduces \sys, an innovative in-memory vector index design optimized for multi-tenant queries. It offers low memory usage, along with high search, insertion, and deletion efficiency. \sys indexes each tenant's vectors with a tenant-specific clustering tree and encodes these trees compactly as sub-trees of a shared clustering tree. Each tenant's clustering tree adapts dynamically to the tenant's unique vector distribution. Our evaluations using two popular datasets demonstrate that \sys matches the search efficiency of individual tenant indexing while keeping memory usage as low as with metadata filtering approaches. Our code is open-sourced at \url{https://github.com/hatsu3/curator}.

\section*{Acknowledgement}
We thank Jun Yang for helpful discussion and feedback. This research was partially supported by NSF grants CNS-2238665 and CNS-2112562, and by gifts from Adobe, Amazon, Meta, and IBM.

\bibliographystyle{ACM-Reference-Format}
\bibliography{99-ref}


\begin{thebibliography}{59}


\ifx \showCODEN    \undefined \def \showCODEN     #1{\unskip}     \fi
\ifx \showDOI      \undefined \def \showDOI       #1{#1}\fi
\ifx \showISBNx    \undefined \def \showISBNx     #1{\unskip}     \fi
\ifx \showISBNxiii \undefined \def \showISBNxiii  #1{\unskip}     \fi
\ifx \showISSN     \undefined \def \showISSN      #1{\unskip}     \fi
\ifx \showLCCN     \undefined \def \showLCCN      #1{\unskip}     \fi
\ifx \shownote     \undefined \def \shownote      #1{#1}          \fi
\ifx \showarticletitle \undefined \def \showarticletitle #1{#1}   \fi
\ifx \showURL      \undefined \def \showURL       {\relax}        \fi
\providecommand\bibfield[2]{#2}
\providecommand\bibinfo[2]{#2}
\providecommand\natexlab[1]{#1}
\providecommand\showeprint[2][]{arXiv:#2}

\bibitem[Andoni et~al\mbox{.}(2015)]%
        {andoni2015practical}
\bibfield{author}{\bibinfo{person}{Alexandr Andoni}, \bibinfo{person}{Piotr Indyk}, \bibinfo{person}{Thijs Laarhoven}, \bibinfo{person}{Ilya Razenshteyn}, {and} \bibinfo{person}{Ludwig Schmidt}.} \bibinfo{year}{2015}\natexlab{}.
\newblock \showarticletitle{Practical and optimal LSH for angular distance}.
\newblock \bibinfo{journal}{\emph{Advances in neural information processing systems}}  \bibinfo{volume}{28} (\bibinfo{year}{2015}).
\newblock


\bibitem[Aulbach et~al\mbox{.}(2008)]%
        {aulbach2008multi}
\bibfield{author}{\bibinfo{person}{Stefan Aulbach}, \bibinfo{person}{Torsten Grust}, \bibinfo{person}{Dean Jacobs}, \bibinfo{person}{Alfons Kemper}, {and} \bibinfo{person}{Jan Rittinger}.} \bibinfo{year}{2008}\natexlab{}.
\newblock \showarticletitle{Multi-tenant databases for software as a service: schema-mapping techniques}. In \bibinfo{booktitle}{\emph{Proceedings of the 2008 ACM SIGMOD international conference on Management of data}}. \bibinfo{pages}{1195--1206}.
\newblock


\bibitem[Aulbach et~al\mbox{.}(2009)]%
        {aulbach2009comparison}
\bibfield{author}{\bibinfo{person}{Stefan Aulbach}, \bibinfo{person}{Dean Jacobs}, \bibinfo{person}{Alfons Kemper}, {and} \bibinfo{person}{Michael Seibold}.} \bibinfo{year}{2009}\natexlab{}.
\newblock \showarticletitle{A comparison of flexible schemas for software as a service}. In \bibinfo{booktitle}{\emph{Proceedings of the 2009 ACM SIGMOD International Conference on Management of data}}. \bibinfo{pages}{881--888}.
\newblock


\bibitem[Aulbach et~al\mbox{.}(2011)]%
        {aulbach2011extensibility}
\bibfield{author}{\bibinfo{person}{Stefan Aulbach}, \bibinfo{person}{Michael Seibold}, \bibinfo{person}{Dean Jacobs}, {and} \bibinfo{person}{Alfons Kemper}.} \bibinfo{year}{2011}\natexlab{}.
\newblock \showarticletitle{Extensibility and data sharing in evolving multi-tenant databases}. In \bibinfo{booktitle}{\emph{2011 IEEE 27th international conference on data engineering}}. IEEE, \bibinfo{pages}{99--110}.
\newblock


\bibitem[Babenko and Lempitsky(2014)]%
        {babenko2014inverted}
\bibfield{author}{\bibinfo{person}{Artem Babenko} {and} \bibinfo{person}{Victor Lempitsky}.} \bibinfo{year}{2014}\natexlab{}.
\newblock \showarticletitle{The inverted multi-index}.
\newblock \bibinfo{journal}{\emph{IEEE transactions on pattern analysis and machine intelligence}} \bibinfo{volume}{37}, \bibinfo{number}{6} (\bibinfo{year}{2014}), \bibinfo{pages}{1247--1260}.
\newblock


\bibitem[Baranchuk et~al\mbox{.}(2018)]%
        {baranchuk2018revisiting}
\bibfield{author}{\bibinfo{person}{Dmitry Baranchuk}, \bibinfo{person}{Artem Babenko}, {and} \bibinfo{person}{Yury Malkov}.} \bibinfo{year}{2018}\natexlab{}.
\newblock \showarticletitle{Revisiting the inverted indices for billion-scale approximate nearest neighbors}. In \bibinfo{booktitle}{\emph{Proceedings of the European Conference on Computer Vision (ECCV)}}. \bibinfo{pages}{202--216}.
\newblock


\bibitem[Bentley(1975)]%
        {bentley1975multidimensional}
\bibfield{author}{\bibinfo{person}{Jon~Louis Bentley}.} \bibinfo{year}{1975}\natexlab{}.
\newblock \showarticletitle{Multidimensional binary search trees used for associative searching}.
\newblock \bibinfo{journal}{\emph{Commun. ACM}} \bibinfo{volume}{18}, \bibinfo{number}{9} (\bibinfo{year}{1975}), \bibinfo{pages}{509--517}.
\newblock


\bibitem[Chandra et~al\mbox{.}(2001)]%
        {chandra2001parallel}
\bibfield{author}{\bibinfo{person}{Rohit Chandra}, \bibinfo{person}{Leo Dagum}, \bibinfo{person}{David Kohr}, \bibinfo{person}{Ramesh Menon}, \bibinfo{person}{Dror Maydan}, {and} \bibinfo{person}{Jeff McDonald}.} \bibinfo{year}{2001}\natexlab{}.
\newblock \bibinfo{booktitle}{\emph{Parallel programming in OpenMP}}.
\newblock \bibinfo{publisher}{Morgan kaufmann}.
\newblock


\bibitem[Chiu et~al\mbox{.}(2019)]%
        {chiu2019learning}
\bibfield{author}{\bibinfo{person}{Chih-Yi Chiu}, \bibinfo{person}{Amorntip Prayoonwong}, {and} \bibinfo{person}{Yin-Chih Liao}.} \bibinfo{year}{2019}\natexlab{}.
\newblock \showarticletitle{Learning to index for nearest neighbor search}.
\newblock \bibinfo{journal}{\emph{IEEE transactions on pattern analysis and machine intelligence}} \bibinfo{volume}{42}, \bibinfo{number}{8} (\bibinfo{year}{2019}), \bibinfo{pages}{1942--1956}.
\newblock


\bibitem[Clement et~al\mbox{.}(2019)]%
        {clement2019usearxiv}
\bibfield{author}{\bibinfo{person}{Colin~B Clement}, \bibinfo{person}{Matthew Bierbaum}, \bibinfo{person}{Kevin~P O'Keeffe}, {and} \bibinfo{person}{Alexander~A Alemi}.} \bibinfo{year}{2019}\natexlab{}.
\newblock \showarticletitle{On the use of arxiv as a dataset}.
\newblock \bibinfo{journal}{\emph{arXiv preprint arXiv:1905.00075}} (\bibinfo{year}{2019}).
\newblock


\bibitem[Covington et~al\mbox{.}(2016)]%
        {covington2016deep}
\bibfield{author}{\bibinfo{person}{Paul Covington}, \bibinfo{person}{Jay Adams}, {and} \bibinfo{person}{Emre Sargin}.} \bibinfo{year}{2016}\natexlab{}.
\newblock \showarticletitle{Deep neural networks for youtube recommendations}. In \bibinfo{booktitle}{\emph{Proceedings of the 10th ACM conference on recommender systems}}. \bibinfo{pages}{191--198}.
\newblock


\bibitem[Dasgupta and Sinha(2013)]%
        {dasgupta2013randomized}
\bibfield{author}{\bibinfo{person}{Sanjoy Dasgupta} {and} \bibinfo{person}{Kaushik Sinha}.} \bibinfo{year}{2013}\natexlab{}.
\newblock \showarticletitle{Randomized partition trees for exact nearest neighbor search}. In \bibinfo{booktitle}{\emph{Conference on learning theory}}. PMLR, \bibinfo{pages}{317--337}.
\newblock


\bibitem[Fu and Cai(2016)]%
        {fu2016efanna}
\bibfield{author}{\bibinfo{person}{Cong Fu} {and} \bibinfo{person}{Deng Cai}.} \bibinfo{year}{2016}\natexlab{}.
\newblock \showarticletitle{Efanna: An extremely fast approximate nearest neighbor search algorithm based on knn graph}.
\newblock \bibinfo{journal}{\emph{arXiv preprint arXiv:1609.07228}} (\bibinfo{year}{2016}).
\newblock


\bibitem[Fu et~al\mbox{.}(2019)]%
        {fu2019nsg}
\bibfield{author}{\bibinfo{person}{Cong Fu}, \bibinfo{person}{Chao Xiang}, \bibinfo{person}{Changxu Wang}, {and} \bibinfo{person}{Deng Cai}.} \bibinfo{year}{2019}\natexlab{}.
\newblock \showarticletitle{Fast approximate nearest neighbor search with the navigating spreading-out graph}.
\newblock \bibinfo{journal}{\emph{Proceedings of the VLDB Endowment}} \bibinfo{volume}{12}, \bibinfo{number}{5} (\bibinfo{year}{2019}), \bibinfo{pages}{461--474}.
\newblock


\bibitem[Gionis et~al\mbox{.}(1999)]%
        {gionis1999similarity}
\bibfield{author}{\bibinfo{person}{Aristides Gionis}, \bibinfo{person}{Piotr Indyk}, \bibinfo{person}{Rajeev Motwani}, {et~al\mbox{.}}} \bibinfo{year}{1999}\natexlab{}.
\newblock \showarticletitle{Similarity search in high dimensions via hashing}. In \bibinfo{booktitle}{\emph{Vldb}}, Vol.~\bibinfo{volume}{99}. \bibinfo{pages}{518--529}.
\newblock


\bibitem[Gollapudi et~al\mbox{.}(2023)]%
        {gollapudi2023filtered}
\bibfield{author}{\bibinfo{person}{Siddharth Gollapudi}, \bibinfo{person}{Neel Karia}, \bibinfo{person}{Varun Sivashankar}, \bibinfo{person}{Ravishankar Krishnaswamy}, \bibinfo{person}{Nikit Begwani}, \bibinfo{person}{Swapnil Raz}, \bibinfo{person}{Yiyong Lin}, \bibinfo{person}{Yin Zhang}, \bibinfo{person}{Neelam Mahapatro}, \bibinfo{person}{Premkumar Srinivasan}, {et~al\mbox{.}}} \bibinfo{year}{2023}\natexlab{}.
\newblock \showarticletitle{Filtered-DiskANN: Graph Algorithms for Approximate Nearest Neighbor Search with Filters}. In \bibinfo{booktitle}{\emph{Proceedings of the ACM Web Conference 2023}}. \bibinfo{pages}{3406--3416}.
\newblock


\bibitem[Harwood and Drummond(2016)]%
        {harwood2016fanng}
\bibfield{author}{\bibinfo{person}{Ben Harwood} {and} \bibinfo{person}{Tom Drummond}.} \bibinfo{year}{2016}\natexlab{}.
\newblock \showarticletitle{Fanng: Fast approximate nearest neighbour graphs}. In \bibinfo{booktitle}{\emph{Proceedings of the IEEE Conference on Computer Vision and Pattern Recognition}}. \bibinfo{pages}{5713--5722}.
\newblock


\bibitem[{Hugging Face}(2021)]%
        {huggingface_allMiniLM-L6-v2}
\bibfield{author}{\bibinfo{person}{{Hugging Face}}.} \bibinfo{year}{2021}\natexlab{}.
\newblock \bibinfo{title}{sentence-transformers/all-MiniLM-L6-v2 - Hugging Face}.
\newblock \bibinfo{howpublished}{\url{https://huggingface.co/sentence-transformers/all-MiniLM-L6-v2}}.
\newblock
\newblock
\shownote{Accessed: 2023-11-26}.


\bibitem[Hui et~al\mbox{.}(2009)]%
        {hui2009supporting}
\bibfield{author}{\bibinfo{person}{Mei Hui}, \bibinfo{person}{Dawei Jiang}, \bibinfo{person}{Guoliang Li}, {and} \bibinfo{person}{Yuan Zhou}.} \bibinfo{year}{2009}\natexlab{}.
\newblock \showarticletitle{Supporting database applications as a service}. In \bibinfo{booktitle}{\emph{2009 IEEE 25th International Conference on Data Engineering}}. IEEE, \bibinfo{pages}{832--843}.
\newblock


\bibitem[Iwasaki and Miyazaki(2018)]%
        {iwasaki2018optimization}
\bibfield{author}{\bibinfo{person}{Masajiro Iwasaki} {and} \bibinfo{person}{Daisuke Miyazaki}.} \bibinfo{year}{2018}\natexlab{}.
\newblock \showarticletitle{Optimization of indexing based on k-nearest neighbor graph for proximity search in high-dimensional data}.
\newblock \bibinfo{journal}{\emph{arXiv preprint arXiv:1810.07355}} (\bibinfo{year}{2018}).
\newblock


\bibitem[Jayaram~Subramanya et~al\mbox{.}(2019)]%
        {jayaram2019diskann}
\bibfield{author}{\bibinfo{person}{Suhas Jayaram~Subramanya}, \bibinfo{person}{Fnu Devvrit}, \bibinfo{person}{Harsha~Vardhan Simhadri}, \bibinfo{person}{Ravishankar Krishnawamy}, {and} \bibinfo{person}{Rohan Kadekodi}.} \bibinfo{year}{2019}\natexlab{}.
\newblock \showarticletitle{Diskann: Fast accurate billion-point nearest neighbor search on a single node}.
\newblock \bibinfo{journal}{\emph{Advances in Neural Information Processing Systems}}  \bibinfo{volume}{32} (\bibinfo{year}{2019}).
\newblock


\bibitem[Jegou et~al\mbox{.}(2010)]%
        {jegou2010product}
\bibfield{author}{\bibinfo{person}{Herve Jegou}, \bibinfo{person}{Matthijs Douze}, {and} \bibinfo{person}{Cordelia Schmid}.} \bibinfo{year}{2010}\natexlab{}.
\newblock \showarticletitle{Product quantization for nearest neighbor search}.
\newblock \bibinfo{journal}{\emph{IEEE transactions on pattern analysis and machine intelligence}} \bibinfo{volume}{33}, \bibinfo{number}{1} (\bibinfo{year}{2010}), \bibinfo{pages}{117--128}.
\newblock


\bibitem[Johnson et~al\mbox{.}(2019)]%
        {johnson2019faiss}
\bibfield{author}{\bibinfo{person}{Jeff Johnson}, \bibinfo{person}{Matthijs Douze}, {and} \bibinfo{person}{Herv{\'e} J{\'e}gou}.} \bibinfo{year}{2019}\natexlab{}.
\newblock \showarticletitle{Billion-scale similarity search with {GPUs}}.
\newblock \bibinfo{journal}{\emph{IEEE Transactions on Big Data}} \bibinfo{volume}{7}, \bibinfo{number}{3} (\bibinfo{year}{2019}), \bibinfo{pages}{535--547}.
\newblock


\bibitem[King(2021)]%
        {King_2021}
\bibfield{author}{\bibinfo{person}{Timothy King}.} \bibinfo{year}{2021}\natexlab{}.
\newblock \bibinfo{title}{80 percent of your data will be unstructured in five years}.
\newblock
\newblock
\urldef\tempurl%
\url{https://solutionsreview.com/data-management/80-percent-of-your-data-will-be-unstructured-in-five-years/}
\showURL{%
\tempurl}


\bibitem[Lamrous and Taileb(2006)]%
        {lamrous2006divisive}
\bibfield{author}{\bibinfo{person}{Sid Lamrous} {and} \bibinfo{person}{Mounira Taileb}.} \bibinfo{year}{2006}\natexlab{}.
\newblock \showarticletitle{Divisive hierarchical k-means}. In \bibinfo{booktitle}{\emph{2006 International Conference on Computational Inteligence for Modelling Control and Automation and International Conference on Intelligent Agents Web Technologies and International Commerce (CIMCA'06)}}. IEEE, \bibinfo{pages}{18--18}.
\newblock


\bibitem[Le and Mikolov(2014)]%
        {le2014distributed}
\bibfield{author}{\bibinfo{person}{Quoc Le} {and} \bibinfo{person}{Tomas Mikolov}.} \bibinfo{year}{2014}\natexlab{}.
\newblock \showarticletitle{Distributed representations of sentences and documents}. In \bibinfo{booktitle}{\emph{International conference on machine learning}}. PMLR, \bibinfo{pages}{1188--1196}.
\newblock


\bibitem[Lewis et~al\mbox{.}(2020)]%
        {lewis2020retrieval}
\bibfield{author}{\bibinfo{person}{Patrick Lewis}, \bibinfo{person}{Ethan Perez}, \bibinfo{person}{Aleksandra Piktus}, \bibinfo{person}{Fabio Petroni}, \bibinfo{person}{Vladimir Karpukhin}, \bibinfo{person}{Naman Goyal}, \bibinfo{person}{Heinrich K{\"u}ttler}, \bibinfo{person}{Mike Lewis}, \bibinfo{person}{Wen-tau Yih}, \bibinfo{person}{Tim Rockt{\"a}schel}, {et~al\mbox{.}}} \bibinfo{year}{2020}\natexlab{}.
\newblock \showarticletitle{Retrieval-augmented generation for knowledge-intensive nlp tasks}.
\newblock \bibinfo{journal}{\emph{Advances in Neural Information Processing Systems}}  \bibinfo{volume}{33} (\bibinfo{year}{2020}), \bibinfo{pages}{9459--9474}.
\newblock


\bibitem[Li et~al\mbox{.}(2006)]%
        {li2006dynamic}
\bibfield{author}{\bibinfo{person}{Feifei Li}, \bibinfo{person}{Marios Hadjieleftheriou}, \bibinfo{person}{George Kollios}, {and} \bibinfo{person}{Leonid Reyzin}.} \bibinfo{year}{2006}\natexlab{}.
\newblock \showarticletitle{Dynamic authenticated index structures for outsourced databases}. In \bibinfo{booktitle}{\emph{Proceedings of the 2006 ACM SIGMOD international conference on Management of data}}. \bibinfo{pages}{121--132}.
\newblock


\bibitem[MacQueen et~al\mbox{.}(1967)]%
        {macqueen1967some}
\bibfield{author}{\bibinfo{person}{James MacQueen} {et~al\mbox{.}}} \bibinfo{year}{1967}\natexlab{}.
\newblock \showarticletitle{Some methods for classification and analysis of multivariate observations}. In \bibinfo{booktitle}{\emph{Proceedings of the fifth Berkeley symposium on mathematical statistics and probability}}, Vol.~\bibinfo{volume}{1}. Oakland, CA, USA, \bibinfo{pages}{281--297}.
\newblock


\bibitem[Malkov and Yashunin(2018)]%
        {malkov2018hnsw}
\bibfield{author}{\bibinfo{person}{Yu~A Malkov} {and} \bibinfo{person}{Dmitry~A Yashunin}.} \bibinfo{year}{2018}\natexlab{}.
\newblock \showarticletitle{Efficient and robust approximate nearest neighbor search using hierarchical navigable small world graphs}.
\newblock \bibinfo{journal}{\emph{IEEE transactions on pattern analysis and machine intelligence}} \bibinfo{volume}{42}, \bibinfo{number}{4} (\bibinfo{year}{2018}), \bibinfo{pages}{824--836}.
\newblock


\bibitem[Muja and Lowe(2009)]%
        {muja2009fast}
\bibfield{author}{\bibinfo{person}{Marius Muja} {and} \bibinfo{person}{David~G Lowe}.} \bibinfo{year}{2009}\natexlab{}.
\newblock \showarticletitle{Fast approximate nearest neighbors with automatic algorithm configuration.}
\newblock \bibinfo{journal}{\emph{VISAPP (1)}} \bibinfo{volume}{2}, \bibinfo{number}{331-340} (\bibinfo{year}{2009}), \bibinfo{pages}{2}.
\newblock


\bibitem[Muja and Lowe(2014)]%
        {muja2014scalable}
\bibfield{author}{\bibinfo{person}{Marius Muja} {and} \bibinfo{person}{David~G Lowe}.} \bibinfo{year}{2014}\natexlab{}.
\newblock \showarticletitle{Scalable nearest neighbor algorithms for high dimensional data}.
\newblock \bibinfo{journal}{\emph{IEEE transactions on pattern analysis and machine intelligence}} \bibinfo{volume}{36}, \bibinfo{number}{11} (\bibinfo{year}{2014}), \bibinfo{pages}{2227--2240}.
\newblock


\bibitem[Narayanan et~al\mbox{.}(2017)]%
        {narayanan2017graph2vec}
\bibfield{author}{\bibinfo{person}{Annamalai Narayanan}, \bibinfo{person}{Mahinthan Chandramohan}, \bibinfo{person}{Rajasekar Venkatesan}, \bibinfo{person}{Lihui Chen}, \bibinfo{person}{Yang Liu}, {and} \bibinfo{person}{Shantanu Jaiswal}.} \bibinfo{year}{2017}\natexlab{}.
\newblock \showarticletitle{graph2vec: Learning distributed representations of graphs}.
\newblock \bibinfo{journal}{\emph{arXiv preprint arXiv:1707.05005}} (\bibinfo{year}{2017}).
\newblock


\bibitem[Naumov et~al\mbox{.}(2019)]%
        {naumov2019deep}
\bibfield{author}{\bibinfo{person}{Maxim Naumov}, \bibinfo{person}{Dheevatsa Mudigere}, \bibinfo{person}{Hao-Jun~Michael Shi}, \bibinfo{person}{Jianyu Huang}, \bibinfo{person}{Narayanan Sundaraman}, \bibinfo{person}{Jongsoo Park}, \bibinfo{person}{Xiaodong Wang}, \bibinfo{person}{Udit Gupta}, \bibinfo{person}{Carole-Jean Wu}, \bibinfo{person}{Alisson~G Azzolini}, {et~al\mbox{.}}} \bibinfo{year}{2019}\natexlab{}.
\newblock \showarticletitle{Deep learning recommendation model for personalization and recommendation systems}.
\newblock \bibinfo{journal}{\emph{arXiv preprint arXiv:1906.00091}} (\bibinfo{year}{2019}).
\newblock


\bibitem[Nayak(2019)]%
        {Nayak_2019}
\bibfield{author}{\bibinfo{person}{Pandu Nayak}.} \bibinfo{year}{2019}\natexlab{}.
\newblock \bibinfo{title}{Understanding searches better than ever before}.
\newblock
\newblock
\urldef\tempurl%
\url{https://blog.google/products/search/search-language-understanding-bert/}
\showURL{%
\tempurl}
\newblock
\shownote{Accessed: 2023-11-26}.


\bibitem[Neyshabur and Srebro(2015)]%
        {neyshabur2015symmetric}
\bibfield{author}{\bibinfo{person}{Behnam Neyshabur} {and} \bibinfo{person}{Nathan Srebro}.} \bibinfo{year}{2015}\natexlab{}.
\newblock \showarticletitle{On symmetric and asymmetric lshs for inner product search}. In \bibinfo{booktitle}{\emph{International Conference on Machine Learning}}. PMLR, \bibinfo{pages}{1926--1934}.
\newblock


\bibitem[Omohundro(1989)]%
        {omohundro1989five}
\bibfield{author}{\bibinfo{person}{Stephen~M Omohundro}.} \bibinfo{year}{1989}\natexlab{}.
\newblock \bibinfo{booktitle}{\emph{Five balltree construction algorithms}}.
\newblock \bibinfo{publisher}{International Computer Science Institute Berkeley}.
\newblock


\bibitem[{OpenAI}(2023)]%
        {chatgpt_retrieval_plugin}
\bibfield{author}{\bibinfo{person}{{OpenAI}}.} \bibinfo{year}{2023}\natexlab{}.
\newblock \bibinfo{title}{ChatGPT Retrieval Plugin}.
\newblock \bibinfo{howpublished}{\url{https://github.com/openai/chatgpt-retrieval-plugin}}.
\newblock
\newblock
\shownote{Accessed: 2023-11-26}.


\bibitem[Partow(2019)]%
        {bloomfilter}
\bibfield{author}{\bibinfo{person}{Arash Partow}.} \bibinfo{year}{2019}\natexlab{}.
\newblock \bibinfo{title}{C++ Bloom Filter Library}.
\newblock \bibinfo{howpublished}{\url{https://github.com/ArashPartow/bloom}}.
\newblock
\newblock
\shownote{Accessed: 2023-11-26}.


\bibitem[{Pinecone}(2023)]%
        {pinecone}
\bibfield{author}{\bibinfo{person}{{Pinecone}}.} \bibinfo{year}{2023}\natexlab{}.
\newblock \bibinfo{title}{Pinecone}.
\newblock \bibinfo{howpublished}{\url{https://www.pinecone.io/}}.
\newblock
\newblock
\shownote{Accessed: 2023-11-26}.


\bibitem[{Qdrant}(2023)]%
        {qdrant}
\bibfield{author}{\bibinfo{person}{{Qdrant}}.} \bibinfo{year}{2023}\natexlab{}.
\newblock \bibinfo{title}{Qdrant}.
\newblock \bibinfo{howpublished}{\url{https://github.com/qdrant/qdrant}}.
\newblock
\newblock
\shownote{Accessed: 2023-11-26}.


\bibitem[Radford et~al\mbox{.}(2021)]%
        {radford2021learning}
\bibfield{author}{\bibinfo{person}{Alec Radford}, \bibinfo{person}{Jong~Wook Kim}, \bibinfo{person}{Chris Hallacy}, \bibinfo{person}{Aditya Ramesh}, \bibinfo{person}{Gabriel Goh}, \bibinfo{person}{Sandhini Agarwal}, \bibinfo{person}{Girish Sastry}, \bibinfo{person}{Amanda Askell}, \bibinfo{person}{Pamela Mishkin}, \bibinfo{person}{Jack Clark}, {et~al\mbox{.}}} \bibinfo{year}{2021}\natexlab{}.
\newblock \showarticletitle{Learning transferable visual models from natural language supervision}. In \bibinfo{booktitle}{\emph{International conference on machine learning}}. PMLR, \bibinfo{pages}{8748--8763}.
\newblock


\bibitem[Shrivastava and Li(2014a)]%
        {shrivastava2014asymmetric}
\bibfield{author}{\bibinfo{person}{Anshumali Shrivastava} {and} \bibinfo{person}{Ping Li}.} \bibinfo{year}{2014}\natexlab{a}.
\newblock \showarticletitle{Asymmetric LSH (ALSH) for sublinear time maximum inner product search (MIPS)}.
\newblock \bibinfo{journal}{\emph{Advances in neural information processing systems}}  \bibinfo{volume}{27} (\bibinfo{year}{2014}).
\newblock


\bibitem[Shrivastava and Li(2014b)]%
        {shrivastava2014improved}
\bibfield{author}{\bibinfo{person}{Anshumali Shrivastava} {and} \bibinfo{person}{Ping Li}.} \bibinfo{year}{2014}\natexlab{b}.
\newblock \showarticletitle{Improved asymmetric locality sensitive hashing (ALSH) for maximum inner product search (MIPS)}.
\newblock \bibinfo{journal}{\emph{arXiv preprint arXiv:1410.5410}} (\bibinfo{year}{2014}).
\newblock


\bibitem[Singh et~al\mbox{.}(2021)]%
        {singh2021freshdiskann}
\bibfield{author}{\bibinfo{person}{Aditi Singh}, \bibinfo{person}{Suhas~Jayaram Subramanya}, \bibinfo{person}{Ravishankar Krishnaswamy}, {and} \bibinfo{person}{Harsha~Vardhan Simhadri}.} \bibinfo{year}{2021}\natexlab{}.
\newblock \showarticletitle{FreshDiskANN: A Fast and Accurate Graph-Based ANN Index for Streaming Similarity Search}.
\newblock \bibinfo{journal}{\emph{arXiv preprint arXiv:2105.09613}} (\bibinfo{year}{2021}).
\newblock


\bibitem[Sion(2005)]%
        {sion2005query}
\bibfield{author}{\bibinfo{person}{Radu Sion}.} \bibinfo{year}{2005}\natexlab{}.
\newblock \showarticletitle{Query execution assurance for outsourced databases}. In \bibinfo{booktitle}{\emph{Proceedings of the 31st international conference on Very large data bases}}. \bibinfo{pages}{601--612}.
\newblock


\bibitem[{Spotify}(2013)]%
        {annoy}
\bibfield{author}{\bibinfo{person}{{Spotify}}.} \bibinfo{year}{2013}\natexlab{}.
\newblock \bibinfo{title}{Annoy (Approximate Nearest Neighbors Oh Yeah)}.
\newblock \bibinfo{howpublished}{\url{https://github.com/spotify/annoy}}.
\newblock
\newblock
\shownote{Accessed: 2023-11-26}.


\bibitem[Sundaram et~al\mbox{.}(2013)]%
        {sundaram2013streaming}
\bibfield{author}{\bibinfo{person}{Narayanan Sundaram}, \bibinfo{person}{Aizana Turmukhametova}, \bibinfo{person}{Nadathur Satish}, \bibinfo{person}{Todd Mostak}, \bibinfo{person}{Piotr Indyk}, \bibinfo{person}{Samuel Madden}, {and} \bibinfo{person}{Pradeep Dubey}.} \bibinfo{year}{2013}\natexlab{}.
\newblock \showarticletitle{Streaming similarity search over one billion tweets using parallel locality-sensitive hashing}.
\newblock \bibinfo{journal}{\emph{Proceedings of the VLDB Endowment}} \bibinfo{volume}{6}, \bibinfo{number}{14} (\bibinfo{year}{2013}), \bibinfo{pages}{1930--1941}.
\newblock


\bibitem[Thomee et~al\mbox{.}(2016)]%
        {thomee2016yfcc100m}
\bibfield{author}{\bibinfo{person}{Bart Thomee}, \bibinfo{person}{David~A Shamma}, \bibinfo{person}{Gerald Friedland}, \bibinfo{person}{Benjamin Elizalde}, \bibinfo{person}{Karl Ni}, \bibinfo{person}{Douglas Poland}, \bibinfo{person}{Damian Borth}, {and} \bibinfo{person}{Li-Jia Li}.} \bibinfo{year}{2016}\natexlab{}.
\newblock \showarticletitle{YFCC100M: The new data in multimedia research}.
\newblock \bibinfo{journal}{\emph{Commun. ACM}} \bibinfo{volume}{59}, \bibinfo{number}{2} (\bibinfo{year}{2016}), \bibinfo{pages}{64--73}.
\newblock


\bibitem[Wang et~al\mbox{.}(2021)]%
        {wang2021milvus}
\bibfield{author}{\bibinfo{person}{Jianguo Wang}, \bibinfo{person}{Xiaomeng Yi}, \bibinfo{person}{Rentong Guo}, \bibinfo{person}{Hai Jin}, \bibinfo{person}{Peng Xu}, \bibinfo{person}{Shengjun Li}, \bibinfo{person}{Xiangyu Wang}, \bibinfo{person}{Xiangzhou Guo}, \bibinfo{person}{Chengming Li}, \bibinfo{person}{Xiaohai Xu}, {et~al\mbox{.}}} \bibinfo{year}{2021}\natexlab{}.
\newblock \showarticletitle{Milvus: A purpose-built vector data management system}. In \bibinfo{booktitle}{\emph{Proceedings of the 2021 International Conference on Management of Data}}. \bibinfo{pages}{2614--2627}.
\newblock


\bibitem[{Weaviate}(2023)]%
        {weaviate}
\bibfield{author}{\bibinfo{person}{{Weaviate}}.} \bibinfo{year}{2023}\natexlab{}.
\newblock \bibinfo{title}{Weaviate}.
\newblock \bibinfo{howpublished}{\url{https://github.com/weaviate/weaviate}}.
\newblock
\newblock
\shownote{Accessed: 2023-11-26}.


\bibitem[Wei et~al\mbox{.}(2020)]%
        {wei2020analyticdb}
\bibfield{author}{\bibinfo{person}{Chuangxian Wei}, \bibinfo{person}{Bin Wu}, \bibinfo{person}{Sheng Wang}, \bibinfo{person}{Renjie Lou}, \bibinfo{person}{Chaoqun Zhan}, \bibinfo{person}{Feifei Li}, {and} \bibinfo{person}{Yuanzhe Cai}.} \bibinfo{year}{2020}\natexlab{}.
\newblock \showarticletitle{Analyticdb-v: A hybrid analytical engine towards query fusion for structured and unstructured data}.
\newblock \bibinfo{journal}{\emph{Proceedings of the VLDB Endowment}} \bibinfo{volume}{13}, \bibinfo{number}{12} (\bibinfo{year}{2020}), \bibinfo{pages}{3152--3165}.
\newblock


\bibitem[Weissman and Bobrowski(2009)]%
        {weissman2009design}
\bibfield{author}{\bibinfo{person}{Craig~D Weissman} {and} \bibinfo{person}{Steve Bobrowski}.} \bibinfo{year}{2009}\natexlab{}.
\newblock \showarticletitle{The design of the force. com multitenant internet application development platform}. In \bibinfo{booktitle}{\emph{Proceedings of the 2009 ACM SIGMOD International Conference on Management of data}}. \bibinfo{pages}{889--896}.
\newblock


\bibitem[Wu et~al\mbox{.}(2022)]%
        {wu2022hqann}
\bibfield{author}{\bibinfo{person}{Wei Wu}, \bibinfo{person}{Junlin He}, \bibinfo{person}{Yu Qiao}, \bibinfo{person}{Guoheng Fu}, \bibinfo{person}{Li Liu}, {and} \bibinfo{person}{Jin Yu}.} \bibinfo{year}{2022}\natexlab{}.
\newblock \showarticletitle{HQANN: Efficient and Robust Similarity Search for Hybrid Queries with Structured and Unstructured Constraints}. In \bibinfo{booktitle}{\emph{Proceedings of the 31st ACM International Conference on Information \& Knowledge Management}}. \bibinfo{pages}{4580--4584}.
\newblock


\bibitem[Xu et~al\mbox{.}(2011)]%
        {xu2011complementary}
\bibfield{author}{\bibinfo{person}{Hao Xu}, \bibinfo{person}{Jingdong Wang}, \bibinfo{person}{Zhu Li}, \bibinfo{person}{Gang Zeng}, \bibinfo{person}{Shipeng Li}, {and} \bibinfo{person}{Nenghai Yu}.} \bibinfo{year}{2011}\natexlab{}.
\newblock \showarticletitle{Complementary hashing for approximate nearest neighbor search}. In \bibinfo{booktitle}{\emph{2011 International Conference on Computer Vision}}. IEEE, \bibinfo{pages}{1631--1638}.
\newblock


\bibitem[Yury et~al\mbox{.}(2018)]%
        {hnswlib}
\bibfield{author}{\bibinfo{person}{Malkov Yury} {et~al\mbox{.}}} \bibinfo{year}{2018}\natexlab{}.
\newblock \bibinfo{title}{hnswlib}.
\newblock \bibinfo{howpublished}{\url{https://github.com/nmslib/hnswlib}}.
\newblock
\newblock
\shownote{Accessed: 2023-11-26}.


\bibitem[Zeng(2022)]%
        {Zeng_2022}
\bibfield{author}{\bibinfo{person}{Belinda Zeng}.} \bibinfo{year}{2022}\natexlab{}.
\newblock \bibinfo{title}{Go beyond the search box: Introducing multisearch}.
\newblock
\newblock
\urldef\tempurl%
\url{https://blog.google/products/search/multisearch/}
\showURL{%
\tempurl}


\bibitem[Zhang et~al\mbox{.}(2019)]%
        {zhang2019deep}
\bibfield{author}{\bibinfo{person}{Shuai Zhang}, \bibinfo{person}{Lina Yao}, \bibinfo{person}{Aixin Sun}, {and} \bibinfo{person}{Yi Tay}.} \bibinfo{year}{2019}\natexlab{}.
\newblock \showarticletitle{Deep learning based recommender system: A survey and new perspectives}.
\newblock \bibinfo{journal}{\emph{ACM computing surveys (CSUR)}} \bibinfo{volume}{52}, \bibinfo{number}{1} (\bibinfo{year}{2019}), \bibinfo{pages}{1--38}.
\newblock


\bibitem[Zhang et~al\mbox{.}(2014)]%
        {zhang2014composite}
\bibfield{author}{\bibinfo{person}{Ting Zhang}, \bibinfo{person}{Chao Du}, {and} \bibinfo{person}{Jingdong Wang}.} \bibinfo{year}{2014}\natexlab{}.
\newblock \showarticletitle{Composite quantization for approximate nearest neighbor search}. In \bibinfo{booktitle}{\emph{International Conference on Machine Learning}}. PMLR, \bibinfo{pages}{838--846}.
\newblock


\end{thebibliography}

\end{document}